\documentclass[aps,prl,reprint,superscriptaddress,showpacs,nofootinbib]{revtex4-1}

\usepackage{hyperref}
\usepackage{amsmath}
\usepackage{amssymb}
\usepackage{graphicx}
\usepackage{color}
\usepackage{subfig}
\usepackage{bm}
\usepackage{gensymb}
\usepackage{bigints}
\usepackage[usenames,dvipsnames,svgnames,table]{xcolor}
\usepackage{pifont}
\usepackage{aas_macros}
\usepackage{orcidlink}
\usepackage{physics}

\usepackage{subfig}
\usepackage{mwe}
\hypersetup{urlcolor=BlueViolet,
	    citecolor=Plum,
	    linkcolor=PineGreen}

\usepackage[labelfont=bf]{caption}

\hypersetup{
    colorlinks=true,
    linkcolor=Blue,
    filecolor=Blue,
    urlcolor=MidnightBlue,
    citecolor=Blue,
   pdftitle={Cooling age}
}

\begin{document}


\title{Twin Stars in General Relativity and Extended Theories of Gravity}


\author{Eva Lope-Oter}
\email[E-mail: ]{mariaevl@ucm.es}
\affiliation{Department of Theoretical Physics \& IPARCOS, Complutense University of Madrid, E-28040, 
Madrid, Spain}

\author{Aneta Wojnar\orcidlink{0000-0002-1545-1483}}
\thanks{Corresponding author}
\email[E-mail: ]{awojnar@ucm.es}
\affiliation{Department of Theoretical Physics \& IPARCOS, Complutense University of Madrid, E-28040, 
Madrid, Spain}

\begin{abstract}
We explore gravity-independent equations of state for neutron stars, particularly focusing on twin stars. Examining four categories, we emphasize their behavior in both General Relativity and Palatini gravity. Additionally, we discuss a subcategory of type I, which, in the context of General Relativity, does not exhibit twin star phenomena, yet demonstrates this phenomenon in modified gravity. Furthermore, we briefly address challenges associated with the negative trace of the energy-momentum tensor, prevalent in both theories.
\end{abstract}

\maketitle

\section{Introduction}

General Relativity (GR) has demonstrated remarkable predictive power, successfully describing diverse phenomena, from the Solar System's dynamics to black hole behavior. It has also received confirmation through the detection of gravitational waves from black hole mergers \cite{abbott2016observation} and neutron star mergers \cite{abbott2017gw170817}. Nevertheless, GR faces limitations in explaining aspects like dark matter in astrophysics \cite{rubin1980rotational}, the accelerated cosmic expansion (dark energy) \cite{huterer1999prospects}, and early cosmological inflation, which remains enigmatic \cite{copeland2006dynamics,nojiri2007introduction,nojiri2017modified,nojiri2011unified,capozziello2008extended,CANTATA:2021ktz}.

Understanding the maximum mass of neutron stars (NS) is pivotal and depends on a provided equation of state (EoS) and theory of gravity. This leads to the "mass-radius (M-R) degeneracy," where a point on the M-R diagram can correspond to various EoSs in GR or Extended Theories of Gravity (ETG) \cite{Olmo:2019flu}.

Initially, the theoretical NS mass limit was around $M \approx 0.6 M_\odot$ in GR \cite{1939PhRv...55..364T}. However, various EoSs have expanded this limit to roughly three times that value \cite{Ozel:2016oaf}. Recent observations have further challenged this, with a NS mass of approximately $M \approx 2.4 M_\odot$ \cite{Romani:2022jhd} and the potential existence of even heavier compact objects \cite{Abbott:2020khf}.

Determining the maximum NS mass is critical for understanding stellar evolution, supernovae, and binary NS mergers. It primarily hinges on the EoS at high densities ($n>3n_s$), while NS radii at $\approx 1.4M_\odot$ depend on the EoS at lower densities ($n<3n_s$) \cite{Lattimer:2004pg}. Achieving a slope $c_s^2=1$ up to five times nuclear density requires strongly interacting relativistic matter, a challenge for dense matter theory and Quantum Chromodynamics (QCD) \cite{McLerran:2018hbz}.

Additionally, high-density cores may introduce extra degrees of freedom, softening the EoS and limiting maximum NS masses, contributing to the "hyperon puzzle" \cite{Blaschke:2018mqw,Baym:2017whm,Chatterjee:2015pua,Annala:2019puf,McLerran:2018hbz,Leonhardt:2019fua}. One way to address these challenges is to work in the framework of ETG.

In many proposals, the generic field equations can be written as
\begin{equation}
    \sigma (G_{\mu\nu} - W_{\mu\nu}) = \kappa^2 T_{\mu\nu}.
\end{equation}
Here, $ {G}_{\mu\nu}$ denotes the Einstein tensor, while the factor $ {\sigma}( {\Psi}^i)$ signifies a coupling to gravity, where $ {\Psi}$ may represent, for instance, curvature invariants or other fields. The tensor $ {W}_{\mu\nu}$ is a symmetric tensor, which can be considered as an additional geometric term or/and fields. Its specific form depends on the theory of gravity under consideration.
Apart from it, depending on a given theory, one can deal with additional field equations, such as for example Klein-Gordon equation in the case of scalar-tensor theories or/and an equation determining connection dynamics, as it happens in metric-affine geometry. Nevertheless, assuming the spherical-symmetric ansatz 
for the metric 
\begin{equation}\label{gen_metric}
   {g}_{\mu\nu} = \text{diag}\left(-     {b}(   {r}),     {a}(   {r}),     {r}^2,    {r}^2 \sin{   {\theta}}\right).
\end{equation}
one can write down\footnote{Together with Bianchi identity $\nabla_\mu G^{\mu\nu}=0$.} a generic Tolman-Oppenheimer-Volkoff equation \cite{Wojnar:2016bzk,Wojnar:2017tmy} 
\begin{align}\label{tov}
  \left(\frac{ {\Pi}}{ {\sigma}}\right)'=&-\frac{G\mathcal{M}}{c^2 {r}^2}\left(\frac{c^2 {Q}}{ {\sigma}}+\frac{ {\Pi}}{ {\sigma}}\right)
  \left(1+\frac{4\pi  {r}^3\frac{ {\Pi}}{ {\sigma}}}{c^2\mathcal{M}}\right) {a}( {r})\nonumber\\
  & -\frac{2 {\sigma}}{\kappa^2  {r}}\left(\frac{ {W}_{\theta\theta}}{ {r}^2}-\frac{ {W}_{ {r} {r}}}{ {a}}\right)
\end{align}
which provides a direct interpretation of the additional terms appearing because of the modifications to the Einstein's theory (see discussion in \cite{Velten:2016bdk,Kozak:2021vbm}). 
The prime $'$ in equation (\ref{tov}) signifies differentiation with respect to the radial coordinate $  {r}$.
The quantities $  {Q}$ and $  {\Pi}$, which appear in the aforementioned TOV equation, are referred to as the effective energy density and pressure, respectively. They are defined as follows:
\begin{eqnarray}\label{def}
  {Q}(  {r}) :=   {\rho}(  {r}) - \frac{  {\sigma}(  {r})  {W}_{tt}(  {r})}{\kappa^2c^2  {b}(  {r})},\\
\label{def2}
  {\Pi}(  {r}) :=   {p}(  {r}) - \frac{  {\sigma}(  {r})  {W}_{  {r}  {r}}(  {r})}{\kappa^2  {a}(  {r})}.
\end{eqnarray}
In these equations, the energy density $  {\rho}$ and pressure $  {p}$ satisfy the barotropic equation of state, $  {p} =   {p}(  {\rho})$, and they appear in the perfect fluid representation of the energy-momentum tensor $  {T}_{\mu\nu}$:
\begin{equation}
  {T}_{\mu\nu} = (  {\rho} +   {p})  {u}_\mu   {u}_\nu +   {p}  {h}_{\mu\nu},
\end{equation}
where $  {u}^\mu$ is a vector field comoving with the fluid, and ${h}_{\mu\nu} = {g}_{\mu\nu} + u_\mu u_\nu$ represents a projection tensor onto the $3$-dimensional hypersurface.
On the other hand, the mass equation is given as 
\begin{equation}\label{mr}
\mathcal{M}(  {r})= \int^{  {r}}_0 4\pi \tilde{r}^2\frac{  {Q}(\tilde{r})}{  {\sigma}(\tilde{r})} d\tilde{r}.
\end{equation}
It also appears in the solution of the static spherical-symmetric metric:
\begin{equation}\label{mod_geo}
   {a}(  {r})=\left( 1-\frac{2G \mathcal{M}(  {r})}{c^2  {r}} \right)^{-1}.
\end{equation}
It is evident that the set of equations described above encompasses the GR case for $W_{\mu\nu} = 0$ and $\sigma = $ const. It is worth noting that this framework also accommodates the consideration of non-ideal fluid scenarios, as the anisotropies and other dissipative characteristics of the fluid can be incorporated into the tensor $W_{\mu\nu} $.

In the following sections, our focus will be on GR and specific proposals featuring couplings, denoted by $\sigma$ in the equations above. In the case of our particular interest, Palatini $f(\mathcal R)$ gravity, this coupling becomes a function of the trace of the energy-momentum tensor and will be further referred to as $\mathcal I_1$. To facilitate a comparison with GR in dense matter scenarios, we construct Equations of State (EoSs) exclusively constrained by hadron physics and fundamental principles, deliberately avoiding astrophysical constraints dependent on GR \cite{LopeOter:2019pcq,Komoltsev:2021jzg}. These less constrained EoSs serve as valuable tools for testing ETG proposals, as demonstrated in \cite{Lope-Oter:2023urz}.

To provide an overview, the next section will detail EoSs and phase transitions. We will initially employ them within the framework of GR and subsequently in Palatini gravity. Additionally, we will discuss a new subcase of the type I phase transition. Finally, we will draw our conclusions. Throughout the paper, we adopt natural units with $G=c=1$ and $\kappa^2=8\pi$, maintaining the metric signature as $(-,+,+,+)$.

\section{Phase transitions in neutron stars and twin stars}\label{seceos}

Before delving into the study of twin star phenomena within theories that modify Einstein's GR, it is crucial to first investigate equations of state, phase transitions, and their interplay with the twin star problem.

While most models for the EoS are primarily hadronic in nature, there has been speculation that at the core of a neutron star, a deconfined quark core may exist (or other exotic matter the core can be made of), giving rise to a higher-density class of compact stars, known as hybrid stars [21–28]. These phase transitions can introduce a discontinuity in the mass-radius relation, resulting in the formation of a second branch of solutions. This second branch invariably contains a hybrid star with the same mass but a smaller radius than at least one neutron star from the original branch. These two stars are commonly referred to as twin stars \cite{Ivanenko:1965dg,Itoh:1970uw,Glendenning:1998ag,Alford:2004pf,Benic:2014jia,Alford:2015dpa,Alford:2017qgh,Christian:2017jni,Christian:2021uhd}, denoting two neutron stars with equal mass but unequal radii.

Here we will consider first-order phase transitions (PT), which involve latent heat but will not use second-order phase transitions, also called "continuous phase transitions" or crossovers. First-order PT is based on the Gibbs thermodynamic equilibrium condition:
\begin{equation}\label{Gibbscondition}
T_H = T_E = 0 \ , \ \ \ \
\mu = \mu_c \ , \ \ \ \
P_H = P_E := P_c\ ,
\end{equation}
where the subindex "H" signifies the hadronic phase, while "E" represents the exotic phase. This condition determines at what critical chemical potential $\mu_c$ the pressure of the two phases will be equal. Therefore, what actually changes during the phase transition is the energy density, displaying a horizontal segment in the ($P,\varepsilon$) plane (see examples in Fig. \ref{fig:3EoS_Twin}).

In this paper, we will examine the behavior of phase transitions in ETG, particularly those that can give rise to twin stars.

To conduct this analysis, we initiated by categorizing twin stars into four distinct groups, as illustrated in \cite{Christian:2017jni}. In this classification, within the mass-radius relationship, the peak of the hadronic branch is denoted as the first maximum $M_1$, and following the unstable branch, the apex of the hybrid branch is labeled as the second maximum $M_2$. Therefore, we deal with I and II categories in which  $M_1 \geq$ 2$M_\odot$, while $M_2 \geq$ 2$M_\odot$ for type I and $M_2 \leq2 M_\odot$ for type II. On the other hand, $M_1 \geq$  $M_\odot$ and $M_1 \leq$  $M_\odot$ for the cases III and IV, respectively, while $M_2 \geq$ 2$M_\odot$ for both types.

The EoS  used to construct twin-stars have been developed using the same methodology outlined in detail in \cite{LopeOter:2019pcq, Lope-Oter:2023urz}. Specifically, for the low-density regime, we rely on chiral Effective Field Theory ($\chi$EFT) calculations for both pure neutron matter and $\beta$-equilibrated matter at N$^3$LO (next-to-next-to-next-to-leading order, equivalent to the fourth order) as detailed in \cite{Drischler:2020hwi, Drischler:2020yad}. This range covers densities from $n = 0.05 - 0.34$ fm$^{-3}$, encompassing densities from the inner crust to the boundary with the inner core of neutron stars.

For extremely high densities ($n > 40n_s$), where the perturbative Quantum Chromodynamics (pQCD) regime applies, we incorporate the partial $N^3$LO results from \cite{Gorda:2021kme}. On the other hand, in the intermediate region between the chiral domain and the pQCD regime, we perform interpolation, connecting the chiral region at $n = 1.5 n_s$ and the perturbative region at the chemical potential $\mu = 2600$ MeV. In this transitional zone, we establish a grid of potential points ($P,\varepsilon$) that represent candidate EoS, employing an interpolation procedure akin to the one developed in \cite{LopeOter:2019pcq}, with a focus on controlling the slope.

 In this study, we present results from the interpolation at $n=1.5n_s$, as a compromise to build soft EoS at low-intermediate energy densities, to meet the tidal constraints from GW170817  ($\Lambda_{1.4}= 190^{+390}_{-120}$) \cite{LIGOScientific:2018cki}, but stiff enough to meet the radius constraints from NICER ($R\geq $10.8 km at 2 $\sigma$ confidence)  \cite{Miller:2019cac,Riley:2019yda,Miller:2021qha,Riley:2021pdl}.

\subsection{Types of phase transitions in GR}

In what follows, we will discuss in more detail the classes of the PT discussed previously.

Firstly, we should analyze the stability of stars. To do so, one studies the behavior of the mass-radius relation with increasing central pressure. Neutron stars are stable until a maximum in the mass-radius relation is reached, with the condition $\partial M/\partial n_c \geq 0$. That is, the stability of a star requires that its mass must increase with increasing central density (or central pressure). Apart from this, one also utilizes the stability criterion of Seidov (the so-called Seidov limit) \cite{Seidov:1971sv} in GR:
\begin{equation}\label{Seidovsjump}
\Delta \varepsilon_{\text{crit}} := \varepsilon_E-\varepsilon_H = \varepsilon_H \left( \frac{1}{2} +\frac{3}{2}\frac{P_H}{\varepsilon_H} \right),
\end{equation}
which indicates the maximum "critical" discontinuity in the energy density (or jump) in a small-core approximation. When $\Delta \varepsilon < \Delta \varepsilon_{\text{crit}}$, a stable connected hybrid branch continues from the hadronic branch. However, if $\Delta \varepsilon > \Delta \varepsilon_{\text{crit}}$, there is no stable connected branch. Instead, a "third family" of neutron stars emerges, known as the disconnected branch \cite{Alford:2015gna}.

However, the Seidov's criterion should not be used in the framework of ETG since it was derived for GR. To overcome this problem
one can also use the specific latent heat $L$ concept \cite{Lope-Oter:2021mjp} as a way to characterise the intensity of a phase transition. This specific latent heat per nucleon is normalized to the unit mass (therefore, it is a pure number in natural units with $c=1$), defined as
\begin{equation}\label{def:Ln}
L:=\frac{\Delta E}{NM_N}\ ,
\end{equation}
where $M_N=940$ MeV is the vacuum neutron mass, while $\Delta E :=  E_E- E_H$ is the difference of energy in both phases, exotic and hadronic, respectively.

To construct twin stars in a given theory of gravity, we have conducted a preliminary analysis to identify suitable EoS that can generate them. These EoS must have a hadronic branch stiff enough to reach a maximum mass above $2M_\odot$, allowing the reproduction of all four types of twin stars, especially types I and II. This initial analysis is not intended to be exhaustive but serves as a guide in selecting an example EoS for our study in ETG.

   \begin{figure}[h]
      \centering      \includegraphics[width=0.50\textwidth]{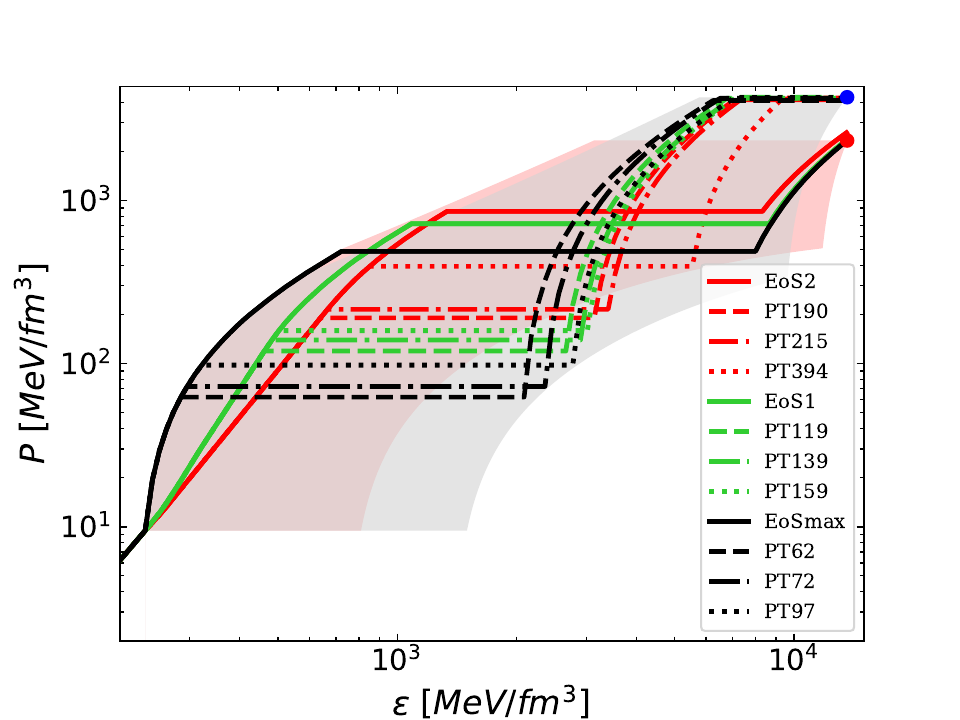}      \caption{{\bf Top:} 3 stiff EoS  to look for twin-stars of types I and II. EoSmax (black) is the most rigid EoS constructed with the maximimum slope from the interpolation point $n=1.5n_s$ allowed by pQCD constraints, EoS1 (green) and EoS2 (red) are EoS of decreasing stiffness, respectively. In the same colours are represented three different PT of each EoS with the maximum jump allowed by pQCD constraints. 
      }
      \label{fig:3EoS_Twin}
  \end{figure}

\begin{table}[h]
    \centering
    \begin{tabular}{cccc}
    \hline
      EoS &  ($\varepsilon_c, P_c$)  & $M_{max}/M_{\odot}$   &$\Delta \varepsilon_{crit}$ \\ 
      \hline \hline
        Max  & (284,62) & 2.12 & 235 \\
          & (295,72) & 2.35 & 256 \\
          & (321,97) & 2.63 & 311 \\
       \hline
       EoS1  & (462,119) & 2.07 & 410 \\
         & (483,139) & 2.19 & 431 \\
         & (502,159) & 2.28 & 450 \\
         & (523,179) & 2.36 & 531 \\
         & (667,319) & 2.63 & 1476 \\
       \hline 
      EoS2  & (640,190) & 2.06 & 606 \\
      & (667,215) & 2.11 & 656 \\
      & (853,394) & 2.35 & 1881 \\
      
    \end{tabular}
    \caption{ EoS points (in MeV/fm$^3$) from  the EoS of Fig. \ref{fig:3EoS_Twin} whose central values yield  similar masses, selected  for phase transitions. The last column collects the maximum jump (in MeV/fm$^3$) allowed by Seidov's limit. }
    \label{tab:PTpoints}
\end{table}

Our starting point has been the three example of stiff EoS of Fig. \ref{fig:3EoS_Twin}: EoSmax (black), EoS1 (green) and EoS2 (red). EoSmax is the most rigid EoS with the maximum slope $c_s^2\lessapprox 1$  at the interpolation point $n_{tr}=1.5n_s$, reaching a maximum mass $M=3.365 M_\odot$; EoS1 (green) and EoS2 (red) are EoS of decreasing stiffness with the same slope as  chiral EoS at $n_{tr}$, but with different acceleration of the slope. EoS1 reaches a maximum mass $M=2.75 M_\odot$ and EoS $M=2.44 M_\odot$. The pQCD constraints force to considerably smooth these EoS at high energy densities, mainly due to the maximum value of the chemical potential assigned at the beginning of the perturbative region ($\mu=2600$ MeV). Therefore, all these EoS are very stiff at inter-region densities, but very soft at high energy densities. In addition, EoS1 and EoS2  satisfy  tidal constraints of GW170817 \cite{LIGOScientific:2018cki}, while EoSmax does not. 
Thus, in all three EoS, the hadronic branch has been built up to a maximum pressure such that the maximum allowed jump and subsequent $c_s^2\leq$1/3 allows to enter pQCD with the assigned limits. All of them show negative trace at $\varepsilon=$348, 523, 694 MeV/fm$^{-3}$ in EoSmax, EoS1 and EoS2, respectively. These values for EoS1 and EoS are in agreement with the results from \cite{Brandes:2023hma} at 9$\%$ and 68$\%$ credible intervals (CIs), respectively,  but not for EoSmax.  

For the phase transitions, we have used those points whose corresponding central values yield similar masses in the three EoS. These values are collected in Table \ref{tab:PTpoints}, showing the different points of EoS1 and EoS2 with similar mass values to EoSmax as well as the maximum  Seidov's jump. Fig. \ref{fig:3EoS_Twin} also displays three PTs of Table \ref{tab:PTpoints}  for each EoS, in the same colours but with different lines, showing in all  the cases the PT with the maximum jump allowed by pQCD, which  requires very high slope, a second PT and a posterior $c_s^2 \leq 1/3$ until entering pQCD.

\subsubsection{Category I and II}

In each of these EoS, we performed phase transitions at the points specified in Table \ref{tab:PTpoints} with various energy jump values to reproduce type I and II twin stars, which pose greater challenges compared to types III and IV.

In general, types I and II necessitate exceptionally high slopes after a PT to attain a stable third branch. Achieving this stability requires meticulous control of the chemical potential to avoid violating the constraints of pQCD, accomplished by smoothing the EoS predictably through prolonged phase transitions. In this context, we have opted for the maximum value $c_s^2=1$ to generate the broadest range of masses for twin stars \cite{Li:2022ivt}. The utilization of this maximal slope implies small negative traces over a substantial energy range, a contentious aspect in QCD \cite{Bedaque:2014sqa,Ecker:2022dlg,Roy:2022nwy,Fujimoto:2022ohj,Marczenko:2022jhl} and gravitation (as discussed in our conclusion section).

   \begin{figure}
      \centering      \includegraphics[width=0.45\textwidth]{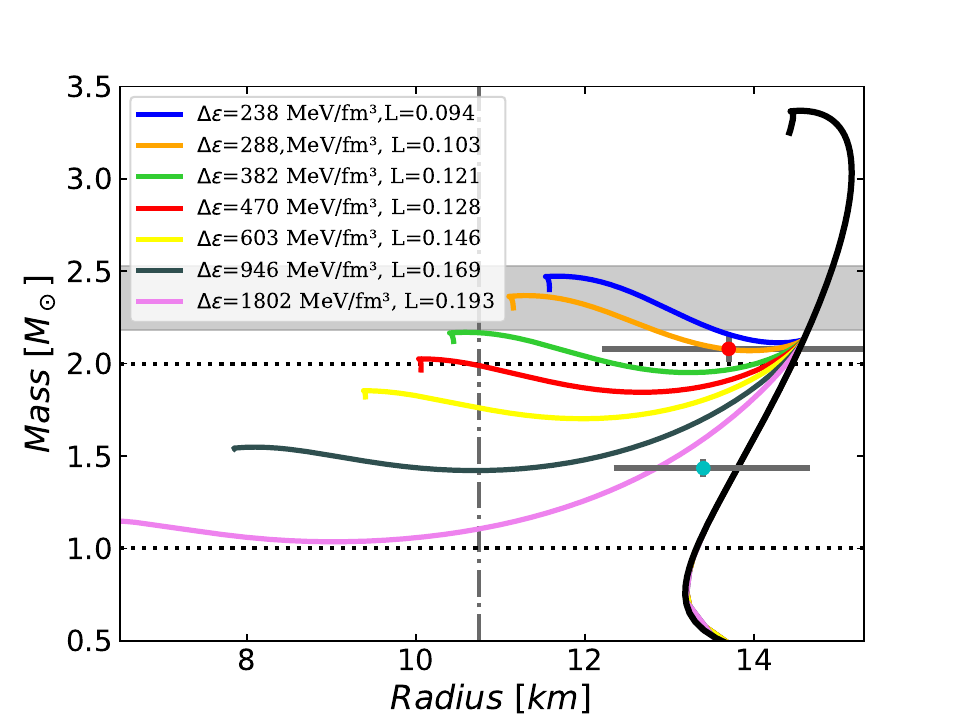}      \includegraphics[width=0.45\textwidth]{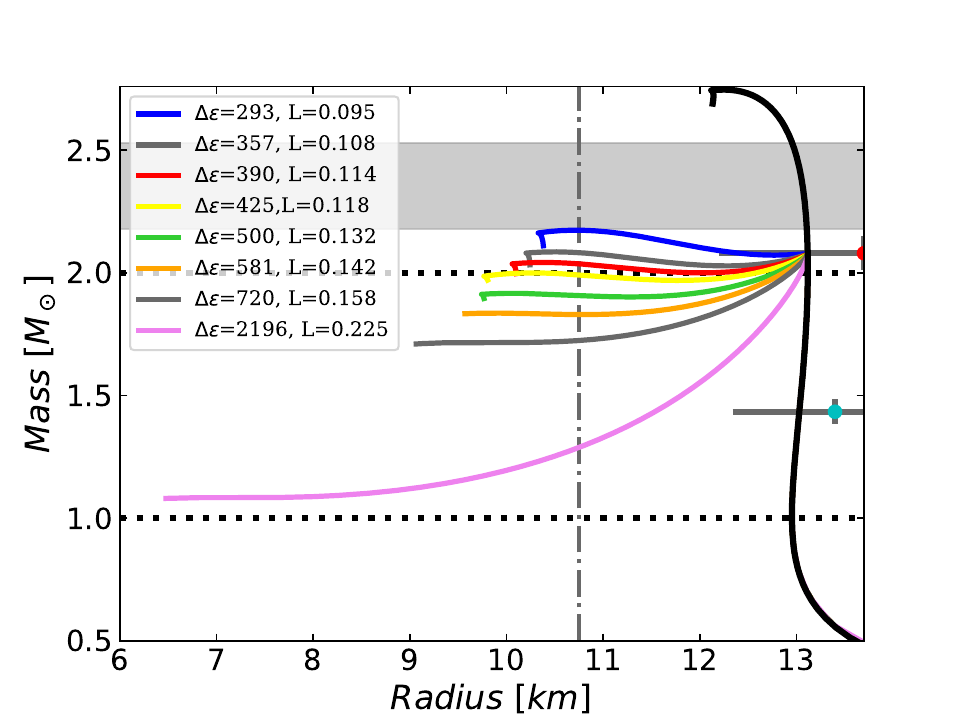}\\     \caption{{\bf Top:} Examples of  type I and II diagrams for PTs in EoSmax (black in Fig) at P=62 MeV/fm$^3$. For this pressure ,twin-stars can be obtained for  any jump $238\leq \Delta \leq 1802$ MeV/fm$^3$. {\bf \textbf{Bottom}:}  Examples of  type I and II diagrams for PTs in EoS1 (green in Fig) at P=119 MeV/fm$^3$, for a similar mass of the pressure at the top. In this case, for jumps $720 \Delta\varepsilon < 2196$ MeV/fm$^3$, we have not found twin-stars type II.}
      \label{fig:PT_typeI_II}
  \end{figure}

Type I twin stars, characterized by $M_1 \geq 2M_\odot$ and $M_2 \geq 2M_\odot$, are realized in three distinct EoS. This classification is achieved by implementing non-extensive jumps ($220\leq \Delta\varepsilon\leq 490$) at each phase transition, while also ensuring that the maximum slope $c_s^2\lessapprox 1$ after the phase transition is maintained in the three EoS (Fig. \ref{fig:3EoS_Twin}). These jumps may fall below or exceed Seidov's limit.

The outcomes for type I twin stars obtained with these EoS are summarized in Table \ref{tab:typeI}, where three dots indicate that other phase transitions at the same point with intermediate jumps also lead to twin star configurations. Notably, no type I twin stars are found for $P\geq$ 190 MeV/fm$^3$ ($n/n_s$=3.7 fm$^{-3}$) in EoS2, $P>$ 159 MeV/fm$^3$ ($n/n_s$=2.97 fm$^{-3}$) in EoS1, and $P>$ 98 MeV/fm$^3$ ($n/n_s$=1.98 fm$^{-3}$) in EoSmax. Therefore, PTs of EoS2 do not generate type I twin stars. The range of the mass gap obtained for this category is $0.01< \Delta M <0.34 M_\odot$, and the latent heat spans $0.094 \leq L \leq 0.149$. We have considered all obtained twins, regardless of the resulting mass gaps. Taking into account $\Delta M > 0.1 M_\odot$ to confirm the existence of twins in a NICER manner \cite{Christian:2021uhd}, some results from Table \ref{tab:typeI} might be rejected in GR.

Type II twin stars ($M_1 \geq 2M_\odot$, $M_2 \geq 1M_\odot$) are obtained with jumps exceeding the Seidov limit. Specifically, in EoSmax, type II twins are obtained for any jump with $524\leq \Delta\varepsilon \leq 2060$ MeV/fm$^3$ for P-72 and $502\leq \Delta\varepsilon \leq 1802$ MeV/fm$^3$ for P-62. In EoS1, type II twins are obtained for P-119 only, for jumps $425 \leq \Delta\varepsilon \leq 720$, plus $\Delta\varepsilon=2196$ MeV/fm$^3$, but no twins are obtained for $720 < \Delta\varepsilon <2196$. In the remaining points of EoSmax and EoS1, as well as in all the points of EoS2, no twins are obtained.

Table \ref{tab:typeII} compiles the main results for type II, where the three dots indicate that other PTs at the same point with intermediate jumps also generate twins. The range of the mass gap obtained for this category is $0.1< \Delta M <1.3 M_\odot$, and the latent heat spans $0.12 \leq L \leq 0.22$. Therefore, considering $\Delta M > 0.1 M_\odot$, all these cases could be confirmed by NICER in GR.

Fig. \ref{fig:PT_typeI_II} illustrates type I and II twin stars for $P=62$ MeV/fm$^3$ ($n/n_s$=1.8 fm$^{-3}$) in EoSmax (top) and $P=119$ MeV/fm$^3$ ($n/n_s$=2.79 fm$^{-3}$) in EoS1 (bottom). In EoSmax, these twin stars are obtained at $P=62$ MeV/fm$^3$ for any jump (between red and blue curves at the top of Fig \ref{fig:PT_typeI_II}) up to the maximum allowed by pQCD constraints. However, in EoS1, at $P=119$ MeV/fm$^3$, only one twin star configuration is produced (violet curve) for $\Delta \varepsilon >720$ MeV/fm$^3$ (gray), representing the maximum allowed jump by pQCD constraints. The rest of the cases tend asymptotically towards a certain value of mass, but this branch never fully stabilizes, not generating twin stars. In both plots, the type I twin stars are represented by blue, gray, and red lines, along with corresponding jump values (in MeV/fm$^3$) and latent heats.

Concerning the Seidov limit, the majority of PTs in EoS1 result in type I twins for jumps $\Delta\varepsilon<\Delta\varepsilon_{crit}$. Notably, in EoSmax, $\Delta\varepsilon> \Delta\varepsilon_{crit}$ for all PTs at $P=62$ and for a few at $PT=72$ MeV/fm$^3$, and therefore, Seidov's limit is not particularly useful for characterizing this twin category.

On a different note, the specific latent heat exhibits a consistent value for all PTs, showing a similar mass difference $\Delta M$ concerning the transition's starting point. For $\Delta M <0.1$, the latent heat $0.09\leq L \leq 0.12$, and for $0.1< \Delta M <1.3$ with $0.13\leq L \leq 0.22$.

\begin{figure}
      \centering
      \includegraphics[width=0.45\textwidth]{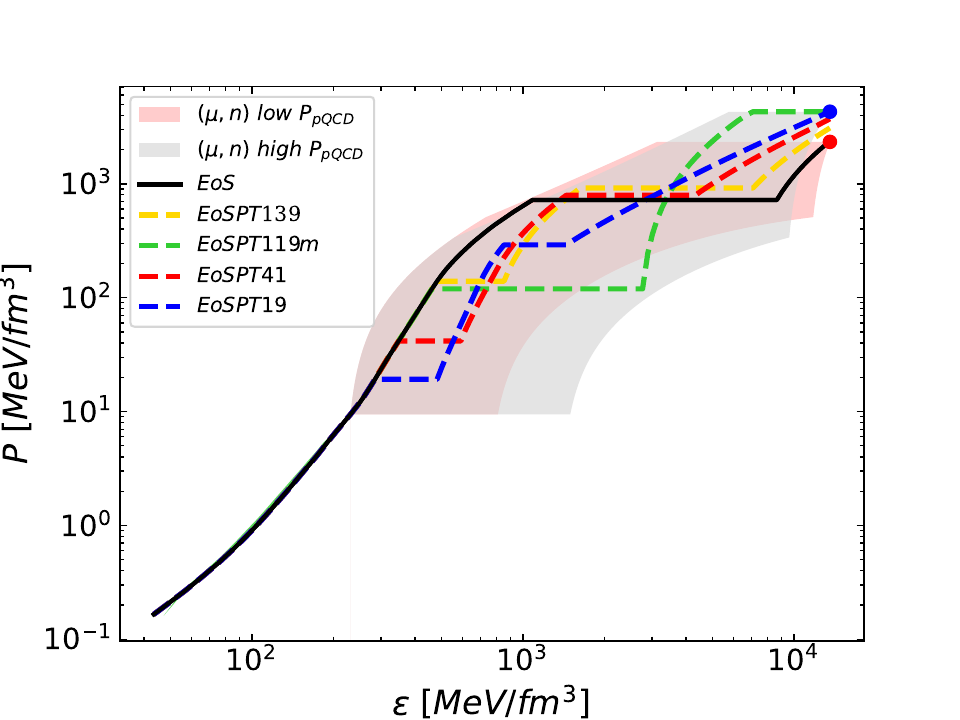}      \includegraphics[width=0.45\textwidth]{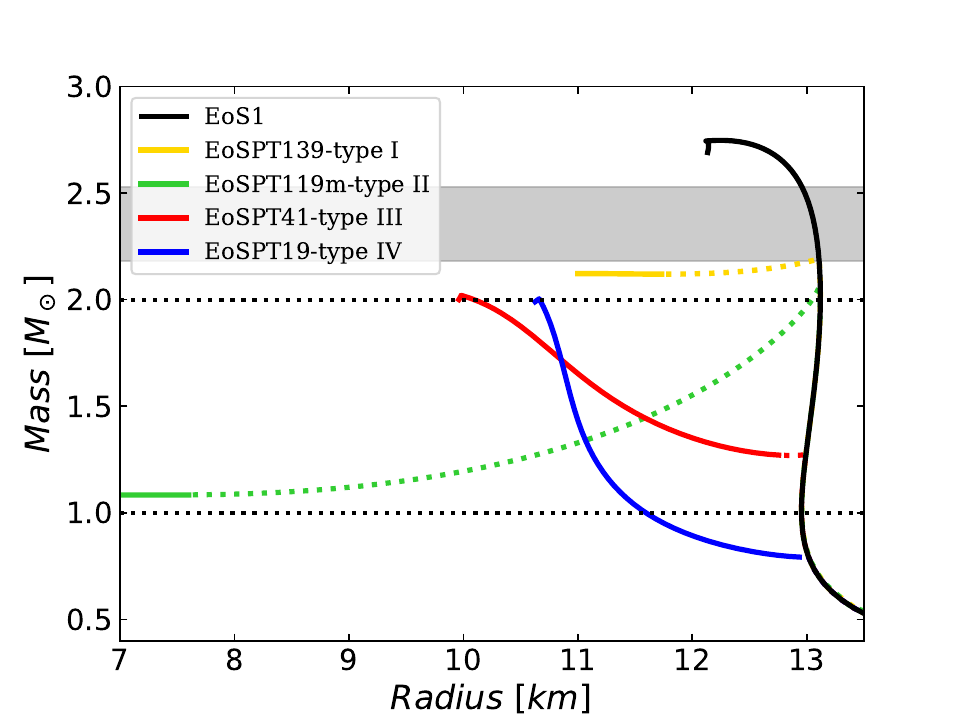}\\     \caption{{\bf Top:} Set of 5 EoSs with $\chi$EFT data from \cite{Drischler:2020yad} and pQCD limits from \cite{Gorda:2021znl} at the matching point $n=1.5n_s$. The black line represents the original EoS1, while the other four correspond to the phase transitions in this EoS at $P=139$ MeV/fm$^3$ (yellow dashed line), $P=119$ MeV/fm$^3$ (green), $P=41$ MeV/fm$^3$ (red), and $P=19$ MeV/fm$^3$ (blue). {\bf \textbf{Bottom}:} M-R diagrams in GR of these EoS indicating the type of twin-star.
      }
      \label{fig:EoS_MvR}
  \end{figure}

For the following analysis we select as an example of type I the EoS from Table \ref{tab:typeI} with PT-139 and a jump with $\Delta\varepsilon=370$ MeV/fm$^3$ (EoSPT139, yellow line in Figure \ref{fig:EoS_MvR}), as a representative case of intermediate latent heat. For type II, we choose the EoS from Table \ref{tab:typeII} with PT-119 and the highest jump with $\Delta\varepsilon=2196$ MeV/fm$^3$  (EoSPT119$_m$, green line in Figure \ref{fig:EoS_MvR}). The main aspects of this Type I and II twin-star are collected in Table \ref{tab:GRalltypes}.

\begin{table}
    \centering
    \begin{tabular}{ccccccc}
    \hline
      $P$ & $\Delta\varepsilon$ (PT) &   $M_1$ & R$_1$& $M_2$ & R$_2$ & L  \\
      MeV/fm$^3$ & MeV/fm$^3$ & $M_\odot$ & km& $M_\odot$ & km & \\ \hline \hline
       EoSmax\\
        97 & 223&2.63&14.99&2.61-2.65&14.42-12.95&0.115 \\
        97 &  246&2.63&14.96&2.58-2.60&13.99-12.90&0.122 \\
        \\
        72 & 228&2.3&14.73&2.29-2.53&14.46-12.06&0.099 \\
         72  & 249 &2.3&14.75&2.27-2.48&14.24-11.87&0.105\\
           ...  & ... & ...&...&...&...&...\\
        72 &  491 &2.3&14.74&1.96-2.03&12.36-10.21&0.149 \\   \\   
        62 &  238 &2.13&14.58&2.11-2.47&14.24-11.67&0.094 \\
           ...  & ... & ...&...&...&...&...\\
        62 &  470 &2.13&14.58&1.84-2.02&12.59-10.03&0.133 \\
        \hline        
      EoS1& \\
        159 &  284 &2.29&13.05&2.28-2.28&12.4-11.9&0.107 \\
        \\
        139 & 304 & 2.19&13.01&2.18,2.20&12.47-11.09&0.105\\
        139 & 370 & 2.19&13.01&2.12-2.122&11.73-11.09&0.12\\
        \\
        119  & 293 & 2.08&13.01&2.07-2.17&12.76-10.72&0.095\\
        119  & 390 & 2.08&13.01&2.0-2.04&12.76-10.72&0.114\\
               \hline        
      EoS2& \\
             \hline        
       190& No  
    \end{tabular}
    \caption{Values of the mass and radius at the peak of the hadronic branch (M$_1$, R$_1$) and the hybrid branch (M$_2$, R$_2$) for PTs at different P values of the EoSs in Figure \ref{fig:3EoS_Twin} and with different jumps $\Delta\varepsilon$ (PT) that generate Type I twin stars. The last column contains the values of the latent heat L associated to each PT. The three dots indicate other PTs at the same point P with all the jumps $\Delta\varepsilon$ included between the previous and the next PT. "No" denotes that there is no PT for the given values of pressure.} 
    \label{tab:typeI}
\end{table}

\begin{table}
    \centering
    \begin{tabular}{ccccccc}
    \hline
      $P$ & $\Delta\varepsilon$ (PT) &   $M_{1max}$ & R$_{M_{1}}$& $M_2$ & R$_2$ & L   \\
      MeV/fm$^3$ & MeV/fm$^3$ & $M_\odot$ & km& $M_\odot$ & km & \\ \hline \hline
      EoSmax\\
        72 & 524 &2.3&14.74&1.92-1.99&12.19-10.19&0.153 \\
         ...  & ... & ...&...&...&...&...\\
        72 & 2060 &2.31&14.75&1.05-1.&8-50-10.21&0.22 \\
        \\
        62 &  502 &2.12&14.58&1.81-1.99&14.24-11.67&0.133 \\
        ...  & ... & ...&...&...&...&...\\
        62 & 1802 &2.125&14.58&1.04-1.10&8.5-6.33&0.194 \\
        \\
      \hline  
      EoS1& \\
        119  & 425 & 2.075&13.01&1.97-1.99&11.76-10.28&0.12\\
        ...  & ... & ...&...&...&...&...\\
       119  & 720 & 2.075&13.1&1.72-1.72&10-9.6&0.158\\
       119  & 2196 & 2.072&13.1&1.083-1.084&7.3-6.9&0.225\\
       \hline
        EoS2& \\                    
       190& No  

    \end{tabular}
    \caption{Values of the mass and radius at the peak of the hadronic branch (M$_1$, R$_1$) and the hybrid branch (M$_2$, R$_2$) for PTs at different P values of the EoSs in Figure \ref{fig:3EoS_Twin} and with different jumps $\Delta\varepsilon$ (PT) that generate Type II twin stars. The last column contains the values of the latent heat L associated to each PT. The three dots indicate other PTs at the same point P with all the jumps $\Delta\varepsilon$ included between the previous and the next PT. "No" denotes that there is no PT for the given values of pressure.} 
    \label{tab:typeII}
\end{table}

\begin{table}
    \centering
    \begin{tabular}{ccccccccc}
    \hline
      & $\Delta\varepsilon_{neg}$&$\Delta \varepsilon$(PT2) & $M_{1}$ & R$_{1}$ & $\varepsilon_{c1}$ & $M_2$ & R$_2$ & $\varepsilon_{c2}$\\
      \hline \hline
       I& 1070-1863&1334-4158 & 2.19 &13.01 & 856 &2.12&11.09& PT2 \\
       \\
       Ia& 1925-6773&3994-13600 & 2.08 &13.12 & 1044 &NA&NA& NA \\
       \\
        \hline        
        II&3836-12500&7080-13600&2.07&13.11&2660&1.08&6.9&5300 \\
        \\
         \hline
        III&948-2355&1446-4158&1.27&12.99&593&2.02&9.98&4159 \\
        \\
         \hline
        IV&819-1043&888-1925&0.79&13.01&463&2.04&10.7&1925 \\        
    \end{tabular}
    \caption{Values of energy density range with negative trace ($\Delta\varepsilon_{neg}$), energy density range for the second PT ($\Delta\varepsilon$ (PT2)), as well as the values of the mass, radius and the corresponding central energy density at the peak of the hadronic branch (M$_1$, R$_1$, $\varepsilon_{c1})$ and the hybrid branch (M$_2$, R$_2$, $\varepsilon_{c2}$) for the EoSs of Figure \ref{fig:EoS_MvR}, generating type I to IV twin stars.} 
    \label{tab:GRalltypes}
\end{table}

\subsubsection{Category III and IV}

Type III and IV twin stars, distinct from type I and II, emerge in all considered EoS for jumps $\Delta\varepsilon > \Delta\varepsilon_{crit}$ with variable jump lengths contingent on the target mass $M_2$. Post PT, diverse slope and slope acceleration choices can be employed to construct the corresponding EoS and achieve type III and IV twin stars.

Both types also require to stiffen the EoS after the PT to achieve $M_2\geq 2M_\odot$, which imply a range of energy density with negative traces ($\Delta\varepsilon_{neg}$). Our goal has been to build EoS that generate these types of twins with $M\geq 2M_\odot$ and with their corresponding radii compatible with NICER ($R\geq $10.8 km). Thus, long jumps, such as $\Delta\varepsilon=350$ MeV/fm$^3$, require the maximum slope to reach that masses and the corresponding radii are far away from the NICER requirements, presenting also very negative traces that, as we already know \cite{Lope-Oter:2023urz}, produce complications in ETG.  
One way to reach a compromise between all these aspects is to reduce the mass gap between both stable branches as much as possible and to grow the EoS after PT with gradual acceleration of the slope.

The type III used here is obtained from EoS1 of Fig. \eqref{fig:EoS_MvR} (red line) by performing a PT at ($\varepsilon, P)= (348,42)$ MeV/fm$^3$ ($n/n_s$=2.19), with an energy density jump $\Delta\varepsilon=242.5$ MeV/fm$^3$ (slightly higher than the Seidov's limit $\Delta\varepsilon_{crit}=238$ MeV/fm$^3$), $L= 0.0485$ and $c_s^2=\{0.42;0.44\}$ at the beginning and end of the PT, respectively. After the PT, the EoS is constructed by a continuous increase of the slope up to $c_s^2 = 0.95$ and then maintaining it (approximately) until a second PT. The main aspects of this Type III twin-star are collected in Table \ref{tab:GRalltypes}.

\begin{figure}
      \centering      \includegraphics[width=0.45\textwidth]{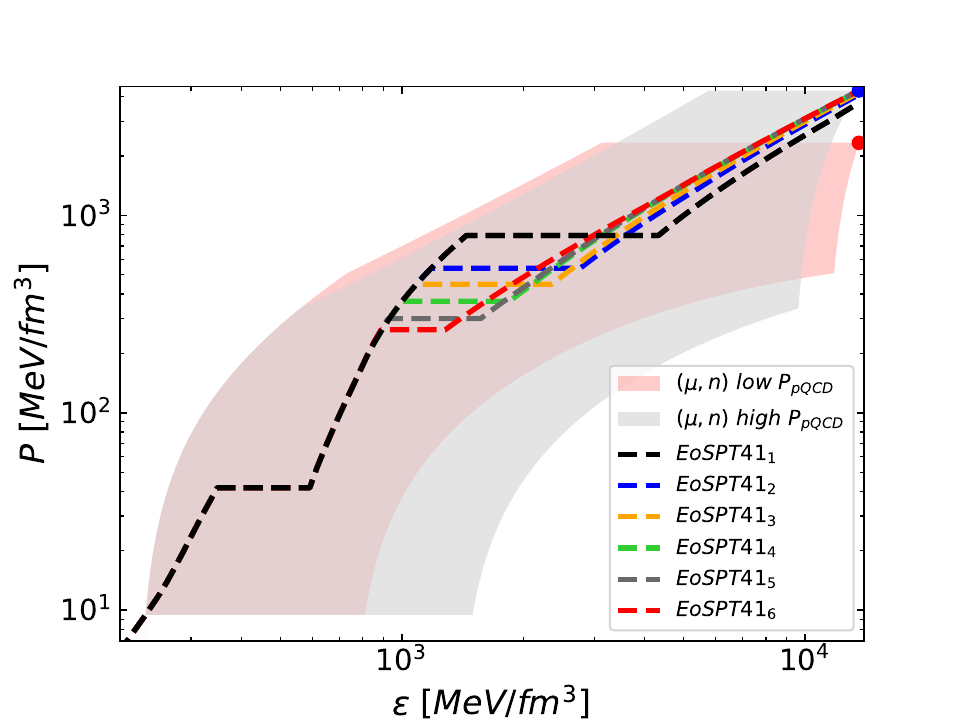}  \includegraphics[width=0.45\textwidth]{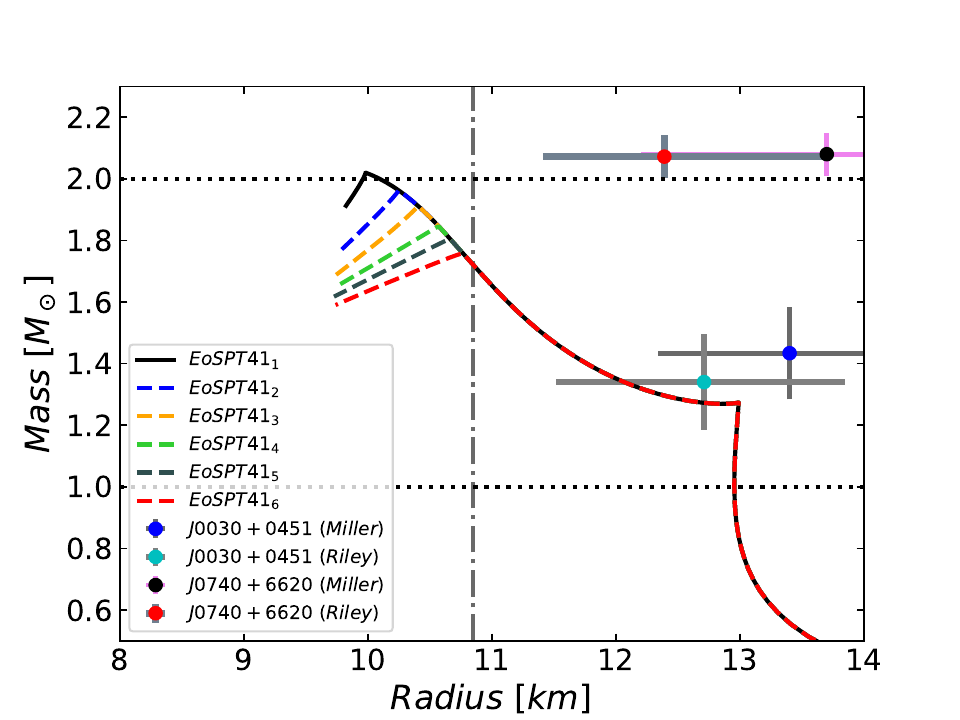}\\     \caption{{\bf Top:} 6 examples of EoS with PT at $P$=41 MeV/fm$^3$ (red in Fig. \ref{fig:EoS_MvR}) with decreasing energy density range with a negative trace. Subscripts 1 to 6 stand for the variation from the highest to the lowest negative trace density range, with 6 being the positive trace EoS. {\bf \textbf{Bottom}:} M-R diagrams in GR of these EoS with the same colors. Black and red circles indicate measurements by NICER of PSR J0740+6620 from \cite{Miller:2021qha} and \cite{Riley:2021pdl}, respectively, with error bars at 1 $\sigma$ confidence; blue and cyan circles indicate measurements by NICER of PSR J0030+0451 from \cite{Miller:2019cac} and \cite{Riley:2019yda}, respectively at 1 sigma confidence.
}
      \label{fig:EoS_PT41}
  \end{figure}

From this initial EoS, we have constructed 5 additional EoS by progressively decreasing $\Delta\varepsilon_{neg}$, as shown in Fig. \ref{fig:EoS_PT41}, performing in each EoS a second PT at lower pressures (lower chemical potential), which involves smaller jumps to continue with $c_s^2 \leq 1/3$ until entering pQCD. In this figure, we denote the initial EoS with the subscript 1 (EoSPT41$_1$), now displayed in black, the subscripts 2 to 5 (blue to gray lines) stand for EoS with decreasing $\Delta\varepsilon_{neg}$, and subscript 6 (red line) for EoS with trace no negative. The values of $\Delta\varepsilon_{neg}$ as well as the minimum value of the trace ($(\varepsilon-3P)_{min}$) for these six EoS are shown in Table \ref{tab:ETG}.
However, decreasing $\Delta\varepsilon_{neg}$ implies a softening of the sequential EoS, progressively reducing the maximum mass of the second stable branch (Fig. \ref{fig:EoS_PT41} at the bottom).

Finally, the type IV (blue line) is obtained from EoS1 by a first PT at  ($\varepsilon, P$) = (285, 19.2) $c_s^2$=0.24, and $\Delta \varepsilon= 178$ MeV/fm$^3$. Moreover, since the jump of the first PT ($\Delta \varepsilon= 178$ MeV/fm$^3$) is slightly higher than the Seidov limit ($\Delta \varepsilon= 171$ MeV/fm$^3$ with a latent heat $L= 0.0258$), there is a small unstable branch ($\Delta \varepsilon= 12$ MeV/fm$^3$) which is not visible in the M-R diagram of Figure \ref{fig:EoS_MvR} (bottom). From the PT end, the EoS is built by an increase in the slope from $c_s^2$=0.26 to $c_s^2 \approx 0.96$, and it is maintained until beginning a second PT at ($\varepsilon, P$) = (888, 357.6) MeV/fm$^3$. In this case, we have chosen the beginning of the second PT so as to keep the trace minimally negative necessary to reach $M=2M_\odot$. The main aspects of this Type IV twin-star are collected in Table \ref{tab:GRalltypes}.

\section{Twin stars in Palatini gravity}\label{secpal}

Our work focuses on Palatini $f(\mathcal{R})$ gravity, a simple metric-affine gravity model \cite{gronwald1997metric,de2010f,baldazzi2022metric}. In this approach, both the metric $g$ and the connection $\hat\Gamma$ are treated as independent variables. Thus, to derive the field equations, one must vary the action for this gravity theory with respect to both these variables. The action takes the form:
\begin{equation}\label{Eq:f(R)}
S=\frac{1}{2 \kappa^2} \int d^4x \sqrt{-g} f(\mathcal{R}) + S_m[g_{\mu\nu},\psi_m].
\end{equation}
Here, $\sqrt{-g}$ is the metric determinant, $f(\mathcal{R})$ is a function of the curvature scalar $\mathcal{R}$, and $S_m$ is the matter action depending only on the metric $g_{\mu\nu}$. Notably, we disregard any dependence of $S_m$ on the connection $\hat\Gamma$ in our study.

The curvature scalar $\mathcal{R}$ is formed from the metric and the Palatini-Ricci curvature tensor $\mathcal{R}_{\mu\nu}(\hat\Gamma)$, with $\mathcal{R} = g^{\mu\nu}\mathcal{R}_{\mu\nu}(\hat\Gamma)$. Ensuring the symmetry of $\mathcal{R}_{\mu\nu}$ is crucial to prevent instabilities \cite{alfonso2017trivial,beltran2019ghosts,jimenez2020instabilities}.

In the case where the function $f$ is linear in $\mathcal{R}$, we recover GR. However, considering $f$ as an arbitrary function of $\mathcal{R}$ leads to a distinct spacetime structure \cite{Allemandi:2004ca,Allemandi:2004wn,olmo2011palatini}, which has implications for stellar descriptions \cite{Olmo:2019flu}.

Variation of Eq. \eqref{Eq:f(R)} with respect to the metric $g_{\mu\nu}$ alone yields the field equations:
\begin{equation}
f'(\mathcal{R})\mathcal{R}_{\mu\nu}-\frac{1}{2}f(\mathcal{R})g_{\mu\nu}=\kappa^2 T_{\mu\nu},
\end{equation}
where $T_{\mu\nu}$ is the energy-momentum tensor of the matter field, calculated as:
\begin{equation}\label{Eq: energytensor}
T_{\mu\nu}=-\frac{2}{\sqrt{-g}}\frac{\delta S_m}{\delta g_{\mu\nu}}.
\end{equation}

The curvature scalar $\mathcal{R}$ can be expressed algebraically as a function of matter fields for a given $f(\mathcal{R})$, as seen in the $g$-trace of the field equations:
\begin{equation}
f'(\mathcal{R})\mathcal{R}-2 f(\mathcal{R})=\kappa^2 T.
\end{equation}

Let us briefly discuss the possible range of the parameter $\beta$ for which we will examine the effects of model on the phase transitions of the matter, and twin stars' properties. Analyzing the weak-field limit shows $|\beta|<2\times 10^{8},\text{m}^2$ \cite{Olmo:2005zr}. Investigating this further, it was shown that Solar System experiments lack the precision to constrain $\beta$ \cite{Toniato:2019rrd}. However, incorporating microphysics, Earth's seismic data limits $|\beta|\lesssim 10^9 ,\text{m}^2$ \cite{Kozak:2023axy,Kozak:2023ruu}.
Moreover, as demonstrated,
in the non-relativistic limit only the quadratic term is significant, with higher-order terms appearing at the sixth order or beyond \cite{Toniato:2019rrd}. The microscopic stability, on the other hand, requires that $\beta >-1.88×\times10^7 \,\text{m}^2$\cite{Wojnar:2023bvv}. These models, like GR, struggle to explain galaxy rotation curves \cite{Hernandez-Arboleda:2022rim,Hernandez-Arboleda:2023abv}, yielding no constraints from galaxy catalogs to date.

In our previous work \cite{Lope-Oter:2023urz}, the neutron star EoS analysis combined with observations tightens it to $|\beta| \lesssim 10^6$ $m^2$. We have firstly reduced the possible values of the parameter by rejecting the values which could provide to the singular behaviour of the TOV and mass equations for the considered EoSs. That is, as discussed in \cite{Kozak:2021vbm} and further elaborated by us in \cite{Lope-Oter:2023urz}, is the EoS dependent property.

\subsection{Stellar structure equations}
We then focus on the specific model $f(\mathcal{R}) = \mathcal R + \beta \mathcal R^2$, known as the Starobinsky or quadratic model. The solution of the field equations in this case resembles GR, implying that any modifications arise from matter fields, parametrized by the single parameter $\beta$.

Subsequently, the Tolman-Oppenheimer-Volkoff (TOV) equation for compact stellar objects modeled as perfect fluids in Palatini gravity are \cite{Wojnar:2016bzk,Wojnar:2017tmy,Kozak:2021vbm}:
\begin{align}\label{tovJ1}
        p' =  \Bigg[&-\frac{G\mathcal{M}(r)}{r^2\mathcal{I}^{1/2}_1}(\varepsilon+p)\left(1 - \frac{2G\mathcal{M}(r)}{r\mathcal{I}^{1/2}_1}\right)^{-1} \nonumber \\
        &\times
   \left(1 + \frac{4\pi\mathcal{I}_1^\frac{3}{2}r^3}{\mathcal{M}(r)}\left(\frac{p}{\mathcal{I}^2_1} + \frac{\mathcal{I}_2}{2\kappa^2}\right)\right)\Bigg]  \\
        &\times \left(\frac{r}{2}\partial_{r} \ln\mathcal{I}_1 + 1 \right)
   +\left(-\varepsilon + 5p\right)\partial_{r} \ln\mathcal{I}_1,\nonumber
\end{align}
 where the functions $\mathcal{I}_1$ and $\mathcal{I}_2$ in the case of a perfect-fluid stress-energy tensor are  ($c^2\rho=\varepsilon$):
\begin{align}
    & \mathcal{I}_1 =  1 + 4\beta \kappa^2 (\varepsilon - 3p) \label{Iota1}, \\
    & \mathcal{I}_2 = \frac{4\beta\kappa^4(\varepsilon - 3p)^2}{\left(1 + 4\beta \kappa^2 (\varepsilon - 3p) \right)^2} \label{Iota2}.
\end{align}
The effective mass function has the following form:
\begin{equation}\label{masaRel} 
\begin{split}
    \mathcal{M}(r) &=    \int^{r}_0 4\pi\tilde{r}^2 \frac{\varepsilon - 2\beta\kappa^2(\varepsilon - 3p)^2}{\left(1 + 4\beta \kappa^2 (\varepsilon - 3p) \right)^{1/2}} \nonumber \\
    &\times\left[1+ \frac{\tilde{r}}{2}\partial_{\tilde{r}}\ln\left(1+4\beta\kappa^2 (\varepsilon -3p)\right)\right]d\tilde{r}.
\end{split}
\end{equation}
These equations are exact and require no approximations, providing a comprehensive description of stellar equilibrium in Palatini gravity. When $\beta=0$, they reduce to the corresponding GR equations, allowing for a direct comparison between the two gravity models using our equations of state.

To facilitate our discussion, we introduce a convenient redefinition of the Starobinsky parameter:
\begin{equation}
\alpha := 2\kappa^2 \beta.
\end{equation}
This adjustment ensures that both $\alpha$ and $\beta$ share the same units (km$^2$), assuming natural units with $c=G=1$. This choice aligns with the units of the Ricci tensor, energy density $\varepsilon$, and pressure $p$, which are expressed in km$^{-2}$.

\subsection{Numerical approach}

Most sufficiently stiff EoS encompass an energy density range where the trace can become negative, resulting in significantly large (small) values of $\mathcal{I}_1$ for positive (negative) values of alpha, depending on the specific alpha value. Furthermore, in cases involving phase transitions of considerable length, particularly for sufficiently small negative alphas, the value of $\mathcal{I}_1$ can reach zero or even turn negative, which contradicts physical interpretation. Therefore, it is essential to acknowledge that the TOV equation \eqref{tovJ1} and the mass equation \eqref{masaRel} become singular when $\mathcal{I}_1=0$. This singularity arises due to the fact that these equations were derived using a conformal transformation, which only encompasses a subset of feasible solutions \cite{Wojnar:2017tmy, Kozak:2021vbm}. Consequently, there exists a singular value of the parameter $\alpha$, denoted as $\alpha_s$, that depends on the trace of the stress-energy tensor:
\begin{equation}\label{singu} 
\alpha_s=-\frac{1}{2(\varepsilon-3P)}.
\end{equation}
The singular value $\alpha_s$ is accountable for the unbounded nature of the stellar mass in \eqref{masaRel} and also for the singular behavior observed in the TOV equation \eqref{tovJ1}.

Hence, in order to address the equations \eqref{tovJ1} and \eqref{masaRel}, it is essential to establish the permissible range of values for the parameter $\alpha$, excluding the singular value $\alpha_s$. Additionally, it should be recognized that the singular value is contingent upon the chosen equation of state (EoS), implying that each energy density can be linked to a distinct $\alpha_s$.

After establishing a range of allowable values for $\alpha > \alpha_s$ (as extensively discussed in \cite{Lope-Oter:2023urz}), the subsequent task is to identify, among these values, those that generate mass-radius diagrams adhering to the constraints set by astrophysical observations, particularly the measured masses and radii of neutron stars. These observations are typically obtained within the framework of GR. However, this approximation is acceptable since it applies in a regime of low or negligible density outside the star. 

In terms of mass constraints, we take into consideration both the lower and upper limits of the mass estimate for PSR J0952-0607 ($2.35 \pm 0.17$ $M_\odot$) as reported in \cite{Romani:2022jhd}. As for radius constraints, we account for measurements acquired by the NICER experiment concerning neutron stars J0740+6620 \cite{Miller:2021qha, Riley:2021pdl} and J0030+0451 \cite{Miller:2019cac,Riley:2019yda}. Consequently, we need to exclude EoS that predict radii $R < 10.8$ km for $M= 2.0 M_\odot$ (at 2$\sigma$ confidence). However, we will use both observations as reference points in all graphs at 1$\sigma$ confidence.  Furthermore, it is noteworthy that some EoSs might not meet the observational constraints in the framework of GR, but could still adhere to them for specific values of $\alpha$ within the acceptable range in Palatini gravity.

Unlike GR, ETG does not tolerate long negative traces ($\Delta\varepsilon_{neg}$) for small negative values of $\alpha$, showing instabilities and divergences for $\alpha<\alpha_{min}$, so that for each minimum value of the trace ($(\varepsilon-3P)_{min}$) there is a corresponding $\alpha_{min}$. All these values  for each EoS are shown in Table \ref{tab:ETG}. Furthermore, the variation of the trace with the energy density is reflected in the behaviour of $\mathcal{I}_1$, which follows the same trend for positive $\alpha$ and the opposite for negative ones. Thus, as the EoS becomes rigid, $\mathcal{I}_1$ decreases/increases according to positive/negative alpha. During phase transitions, the behavior changes, reaching a maximum at the end of the PT for positive alphas and a minimum for negative ones. Consequently, the singular alpha value is minimum at the end of the PT for positive alphas and  maximum for negative alphas. 

The outcomes derived from our numerical analysis, implemented by the Runge-Kutta method, for the employed EoS' are summarized as follows. Fig. \ref{fig:MvREoSPT139}, \ref{fig:MvREoSPT119max}, \ref{fig:MvREoSPT41_1}, and \ref{fig:MvREoSPT19} show the Mass-Radius, Mass-central density $\varepsilon_c$, $I_1-\varepsilon_c$, and Radius-$\varepsilon_c$ relations for Categories I, II, III, and IV, respectively, for GR and Palatini gravity. Now we will discuss each case in detail.

In all the plots, we represent lower and upper limits of the mass estimate for PSR J0952-0607  by grayish shaded area. The lower limit of the radius  from NICER measurements at 2$\sigma$ is indicated by  a vertical dash-dotted line. Moreover, the graphs corresponding a GR and positive alphas are represented by black and coloured lines, respectively, while negative alphas are represented by coloured dashed lines.  

\subsubsection{Category I}
   \begin{figure}
      \centering      \includegraphics[width=0.45\textwidth]{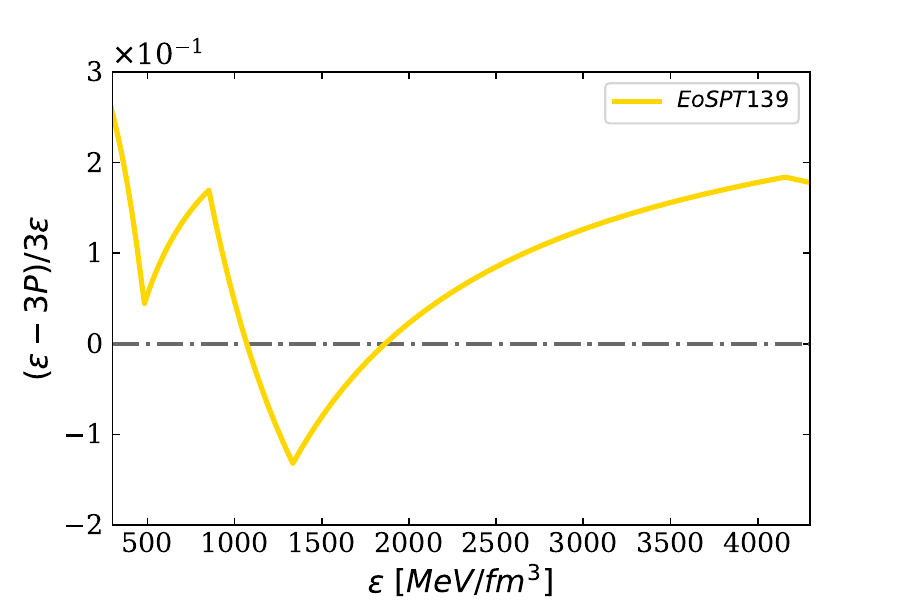}  \includegraphics[width=0.45\textwidth]{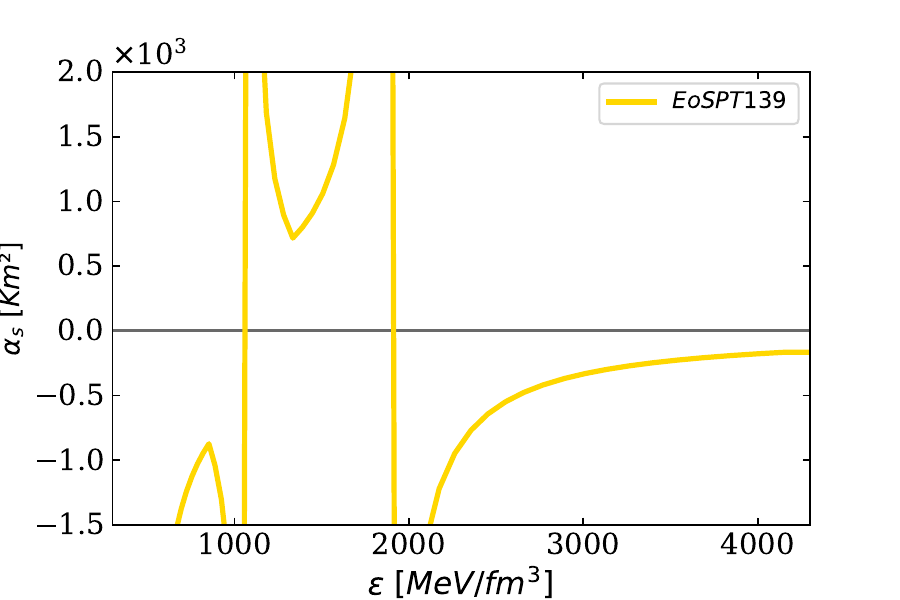}\\     \caption{{\bf Top:} Trace anomaly ($\varepsilon-3P)/3\varepsilon$ as a function of energy density for
the EoSPT139 (yellow curve in Figure \ref{fig:EoS_MvR}). {\bf \textbf{Bottom}:} Values of the singular parameter $\alpha_s$ for the same EoS, showing two singularities for the two null trace points as well as a maximum value of the negative alpha $\alpha_s=-165$ km$^2$ at the end of the second PT, after which it increases slightly until the pQCD regime.   
}
      \label{fig:AnomalyPT139}
  \end{figure}

We exemplify our approach using the EoSPT139, represented by the yellow curve in Figure \ref{fig:EoS_MvR} with a PT at ($483,139.6$) and a jump $\Delta\varepsilon=370$. 

The EoS exhibits a large negative trace range  that can be seen at the top of Figure \ref{fig:AnomalyPT139}. This plot displays the the so-called trace anomaly, or normalized trace $(\varepsilon-3P)/3\varepsilon$), as a function of the energy density, showing a increase of the trace during both PTs, especially for the long second PT, as well as a big decrease afterward during the stiffening of the EoS, taking negative values and reaching the minimum of the trace at the beginning of the second PT. The plot a the bottom of Figure \ref{fig:AnomalyPT139} displays the values of the singular parameter $\alpha_s$ for the same range of energy density, showing the opposite behaviour of the trace, with two singularities for the two null trace points as well as a large increase during the phase transition, especially in the second PT, after which $\alpha_s$ grows slightly, remaining almost constant until the pQCD regime. 

Nevertheless, Palatini gravity disallows for this EoS $\alpha < \alpha_{min}=-85$ km$^2$ that yields instabilities in the range of negative traces $\Delta\varepsilon_{neg}$ until the beginning of the second PT ($\varepsilon_{min}$), where the trace reaches its minimum value.  However, the maximum mass for $\alpha = -85$ km$^2$ is  $M_{max}=2.35 M_\odot$ and in order to achieve  $M_{max}=2.52 M_\odot$ we need $\alpha=-155$ km$^2$, which yields instabilities at the beginning of the range of negative trace ($\Delta\varepsilon_{neg}=15$ MeV/fm$^3$).  

\begin{figure*}[p]
  \centering
  \advance\leftskip-0.7cm
 \subfloat{\includegraphics[width=.48\linewidth]{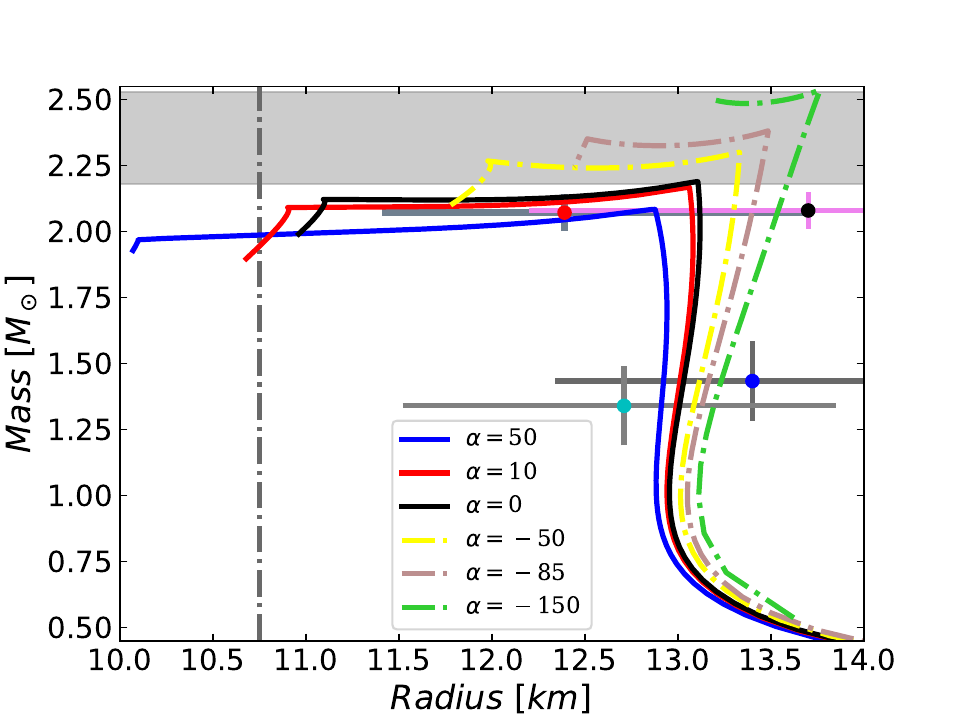}\quad\includegraphics[width=.54\linewidth]{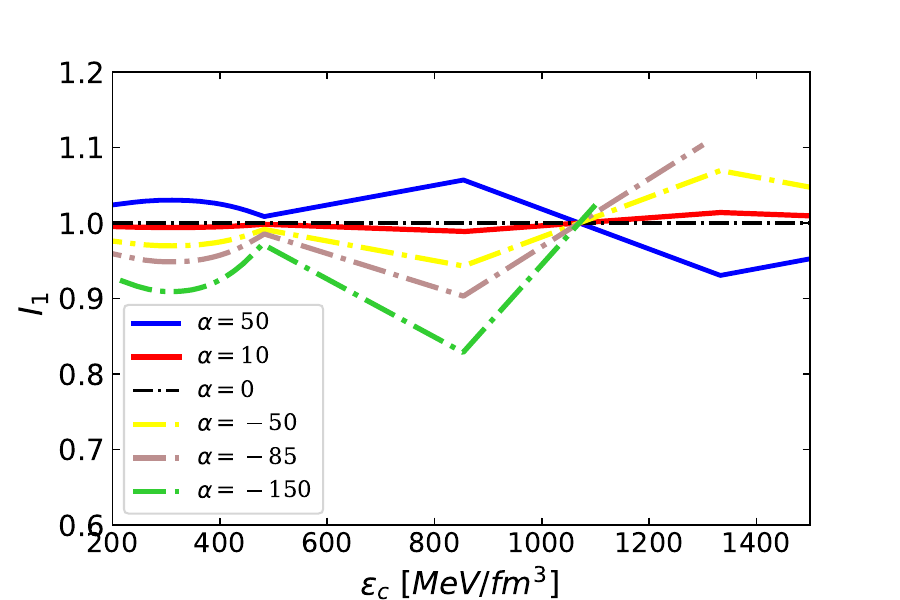}}\\[\baselineskip]
  \subfloat{\includegraphics[width=.515\linewidth]{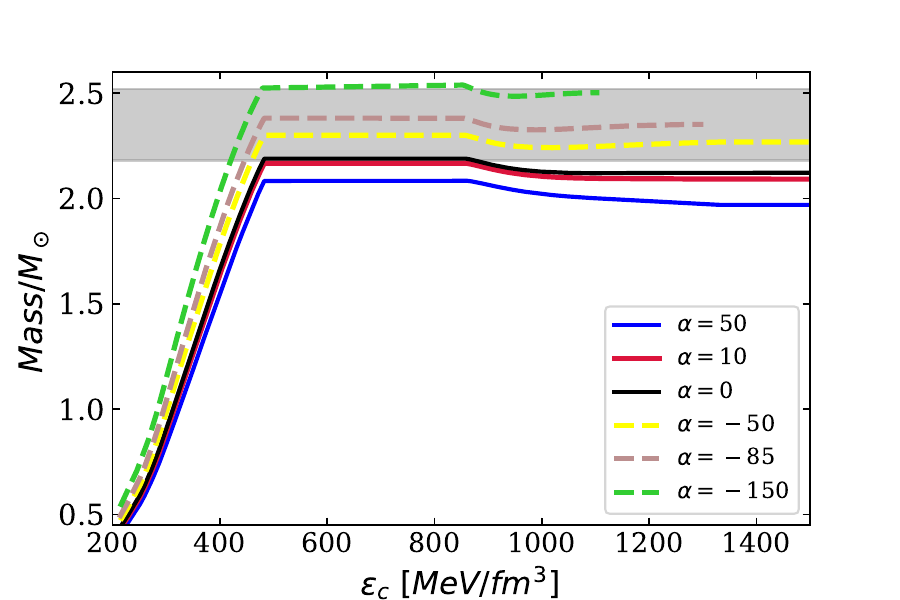}\quad\includegraphics[width=.515\linewidth]{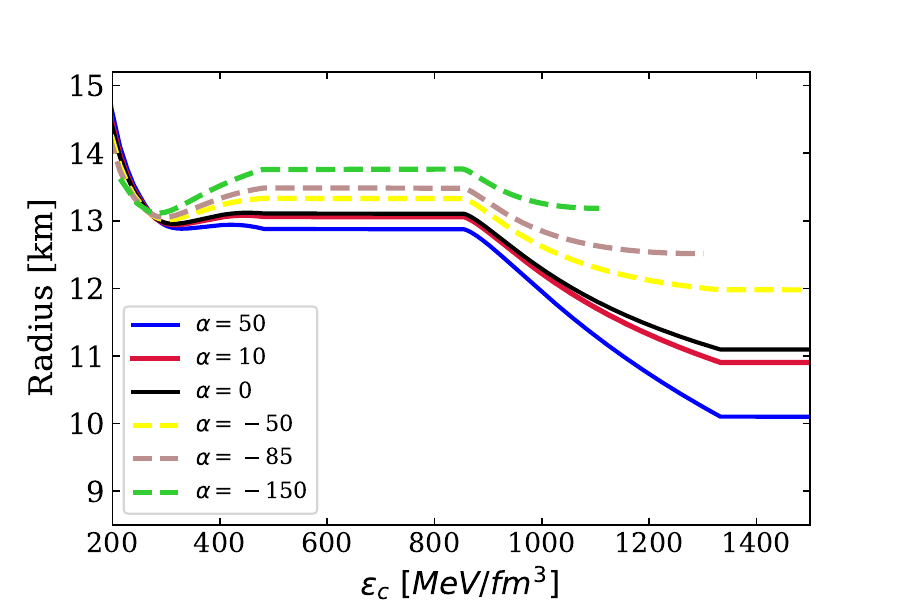}}
\caption{{\bf Top:} Mass-Radius diagrams (left) in GR (black solid line) and ETG (coloured lines) for the  EoSPT139 (yellow)  of Figure \ref{fig:EoS_MvR} as an example of Category I). The grayish shaded area contains the  lower and  upper mass bound (2.17 and 2.52 $M_\odot$, respectively) and the vertical dash-dotted line indicates the lower radius bound (10.8 km). Black and red circles indicate  measurements by  NICER of PSR J0740+6620 from \cite{Miller:2021qha} and \cite{Riley:2021pdl}, respectively, with error bars at 1 sigma confidence; blue and cyan circles indicate measurements by  NICER of PSR J0030+0451 from  \cite{Miller:2019cac} and \cite{Riley:2019yda}, respectively at 1 sigma confidence.  On the right, $\mathcal{I}_1$ against the central energy density for different values of $\alpha$. {\bf Bottom:}  Mass (left) and Radius (right) against the central energy density for green EoS for different values of $\alpha$. }    
    \label{fig:MvREoSPT139}
\end{figure*}

A summary of the results for different values of  Palatini parameter is shown in Figure  \ref{fig:MvREoSPT139}. In ETG, negative values of this parameter ($-300\leq\alpha <0 $) also generate  type I twin-stars for this EoS, but with a larger mass gap, while positive $\alpha\geq 10$ km$^2$ do not produce twins (see top right Figure \ref{fig:MvREoSPT139}). To reach the lower mass bound of $2.18 M_\odot$,  a small positive value of Palatini gravity is  needed ($\alpha=5$ km$^2$) and these results are stable and does not present any divergence. The M-R diagram in ETG at the top of Figure \ref{fig:MvREoSPT139} also shows a second rising branch, with a mass gap similar to GR, but with a lower radius difference, decreasing as alpha decreases. 

The plot at the top right of Figure \ref{fig:MvREoSPT139} shows $\mathcal{I}_1$ values until the point where the code shows instabilities or starts to fail. The bottom left plot of Figure \ref{fig:MvREoSPT139} displays mass against energy density, showing  for negative $\alpha$ a higher mass growth up to the PT than in GR, and a slightly higher mass decrease after PT. Conversely, for positive alphas, the behavior is opposite. The bottom right plot shows radius against energy density, indicating a smaller decrease in radius than in GR until PT for negative alphas, but a smaller decrease after PT. This plot also illustrates correlations between $R$ and $\mathcal{I}_1$, decreasing as $\mathcal{I}_1$ increases while the EoS is stiffening, but its decrease slows down as the slope of $\mathcal{I}_1$ increases (with with decreasing $\alpha$. Moreover, it is worth noting the large decrease of $\mathcal{I}_1$ during the phase transition.

In summary, the  range of $\alpha$ dictated by this EoS  to reach the bounds of grey shade area is $\ -150 \leq \alpha < 5$ km$^2$, showing twin-stars in all this range.

\subsubsection{Category Ia}
   \begin{figure}
      \centering      \includegraphics[width=0.45\textwidth]{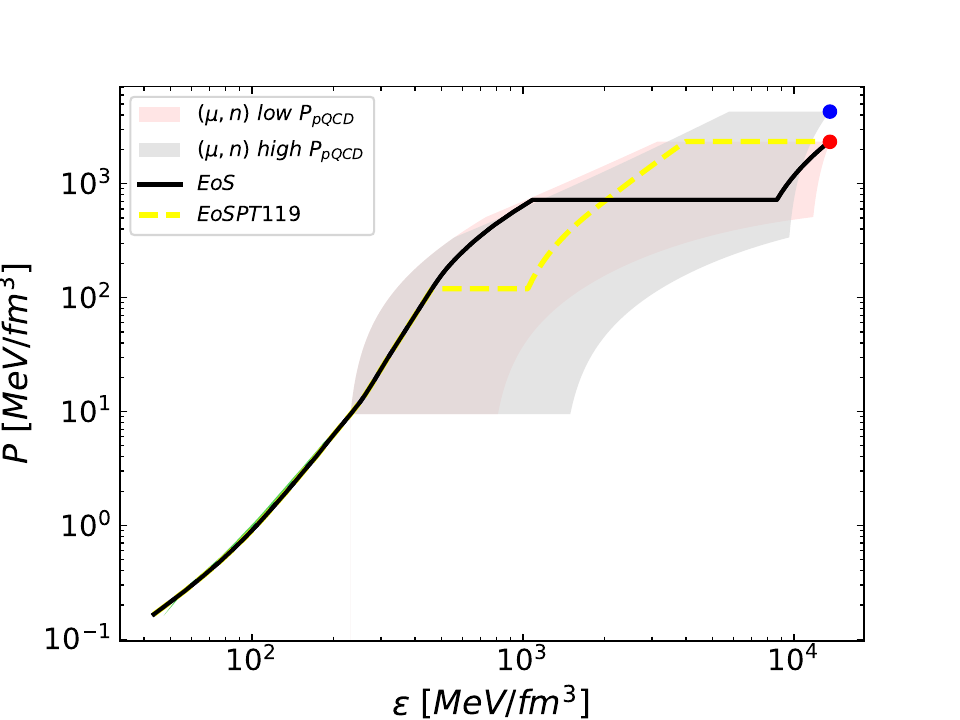}      \includegraphics[width=0.45\textwidth]{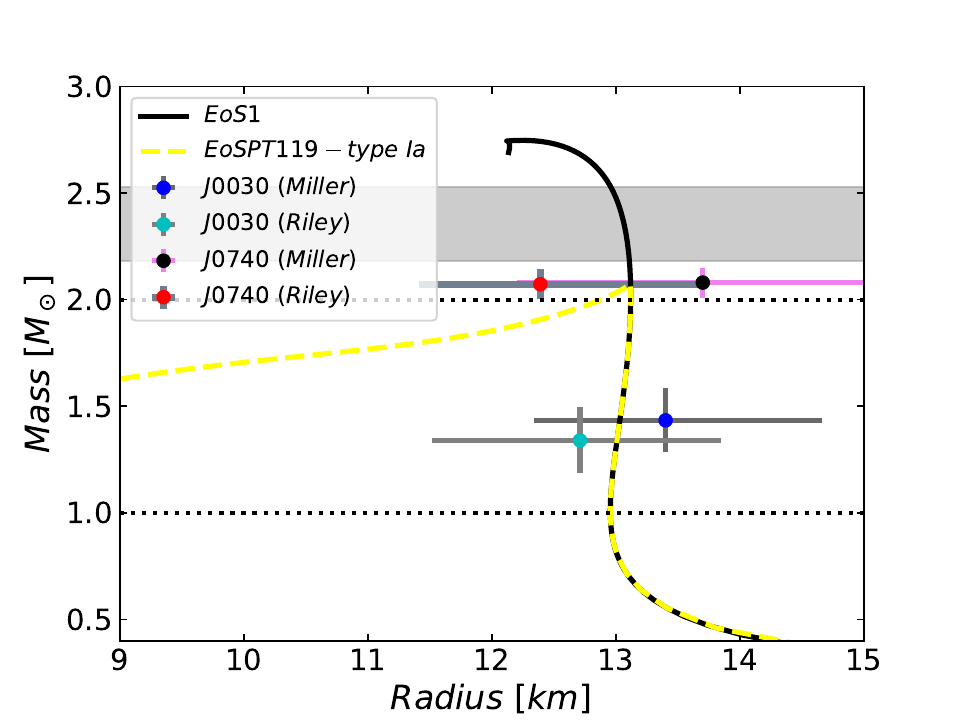}\\     \caption{{\bf Top:} Example of EoSPT119 no twin-star in GR, but twin-star in ETG. {\bf \textbf{Bottom}:} M-R diagram in GR of this EoS.
      }
      \label{fig:EoS_MvR_typeIa}
  \end{figure}

We present in Figure \ref{fig:EoS_MvR_typeIa} an example of an EoS that does not yield a twin star in GR but generates a twin-star category I in ETG. It is represented by the EoSPT119, corresponding to the green curve in Figure \ref{fig:EoS_MvR}, but with a smaller jump at the first PT ($\Delta\varepsilon=581$ MeV/fm$^3$). After this PT, the EoS accelerates its slope from 0.6 up to 0.93 to reach pQCD values, followed by a second PT.

In GR, this EoS reaches only a peak of $M_1= 2.07M_\odot$ and $R=13.1$ km just at the end of the first PT (see Table \ref{tab:GRalltypes}), but becomes unstable afterward. The EoS also exhibits a substantial range of negative trace that is shown in the yellow curve at the top of Figure \ref{fig:AnomalyPT119}. The plot at the bottom of this figure displays the values of the singular parameter, showing that its minimum value of the positive ones $\alpha_s= 125$ km$^2$  is reached at the beginning of the second PT, while maximum value of the  negative ones ($\alpha_s=-57$ km$^2$) is reached at the end of the second PT onset, far away from central values.

\begin{figure*} 
  \centering
  \advance\leftskip-0.7cm
 \subfloat{\includegraphics[width=.48\linewidth]{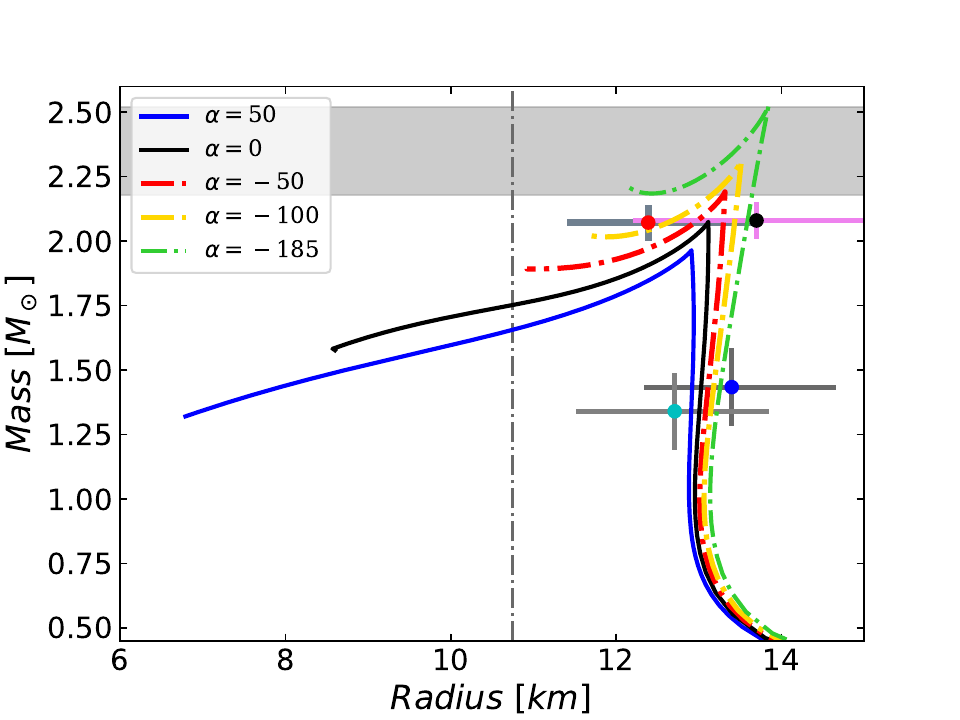}}\quad\includegraphics[width=.54\linewidth]{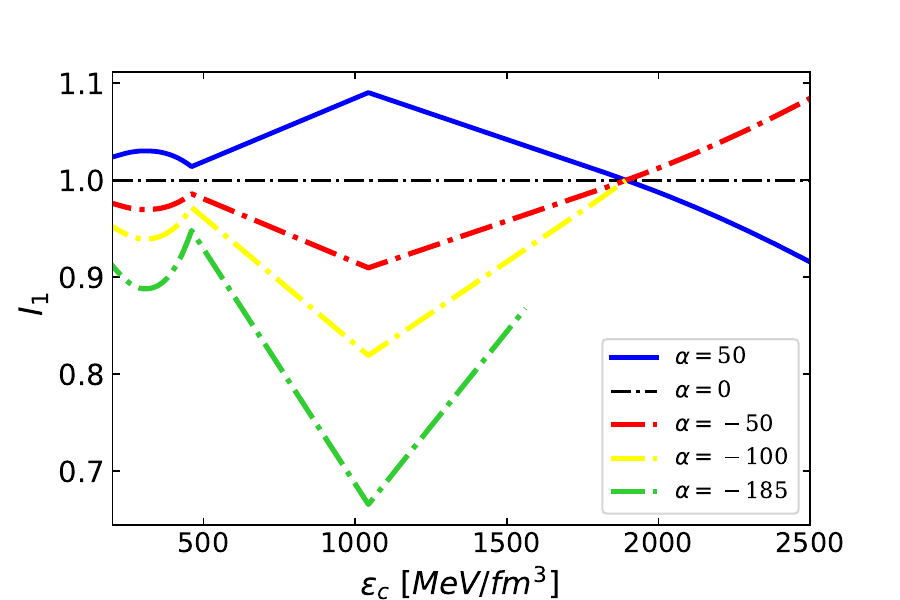}\\[\baselineskip]
  \subfloat{\includegraphics[width=.515\linewidth]{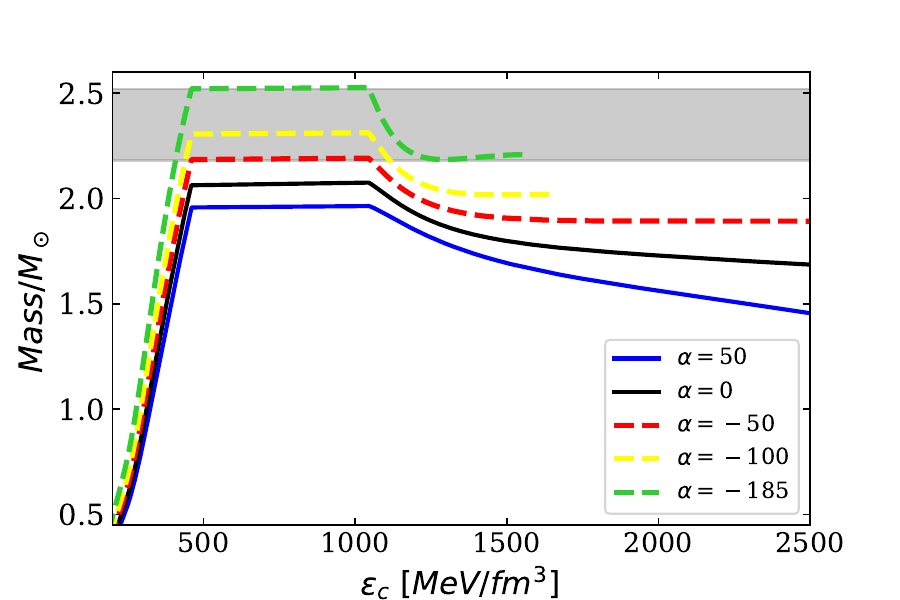}\quad\includegraphics[width=.515\linewidth]{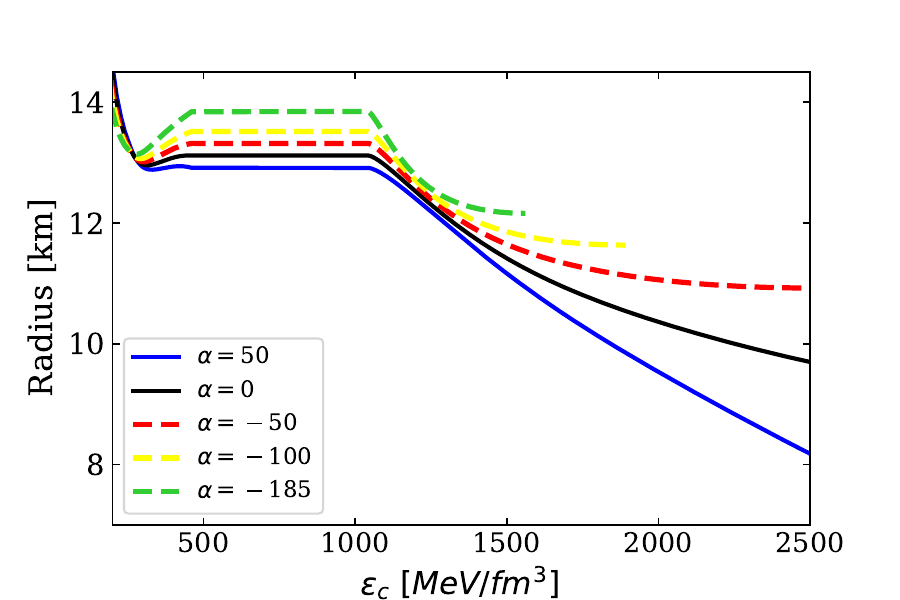}}

\caption{{\bf Top:} Mass-Radius diagrams (left) in GR (black solid line) and ETG (coloured lines) for the  EoSPT119 (yellow)  of Figure \ref{fig:EoS_MvR}, but with a smaller jump, as an example of Category Ia. The grayish shaded area contains the  lower and  upper mass bound (2.17 and 2.52 $M_\odot$, respectively) and the vertical dash-dotted line indicates the lower radius bound (10.8 km). Black and red circles indicate  measurements by  NICER of PSR J0740+6620 from \cite{Miller:2021qha} and \cite{Riley:2021pdl}, respectively, with error bars at 1 sigma confidence; blue and cyan circles indicate measurements by  NICER of PSR J0030+0451 from  \cite{Miller:2019cac} and \cite{Riley:2019yda}, respectively at 1 sigma confidence. On the right, $\mathcal{I}_1$ against the central energy density for different values of $\alpha$. {\bf Bottom:}  Mass against the central energy density for different values of $\alpha$. }    
    \label{fig:MvREoSPT119v4}
\end{figure*}

In ETG, for $\alpha<0$, Type I twins are obtained. We denominate Type Ia to this kind of twin-star. 
The minimum value $\alpha_{min}= -30$km$^2$ associated to the minimum trace is obtained at energy densities $\varepsilon_{min}$ far away from central values. The lower bound of the  mass $M_1=2.19 M_\odot$  is reached  for $\alpha=-50$ km$^2$  at the end of the first PT with its corresponding radius  $R=13.32$ km, showing tolerance to negative trace for $\Delta\varepsilon_{neg}=576$ MeV/fm$^3$. To reach $M=2.52 M_\odot$ requires $\alpha=-185$ km$^2$, for which the code yield instabilities after a range of trace negative $\Delta\varepsilon_{neg}=364$ MeV/fm$^3$.  All these results are shown in Fig.\ref{fig:MvREoSPT119v4} (left), while the right plot displays mass against energy density, where one can see that positive $\alpha$ does not generate twins as in the case of GR.

\subsubsection{Category II}

   \begin{figure}
      \centering      \includegraphics[width=0.45\textwidth]{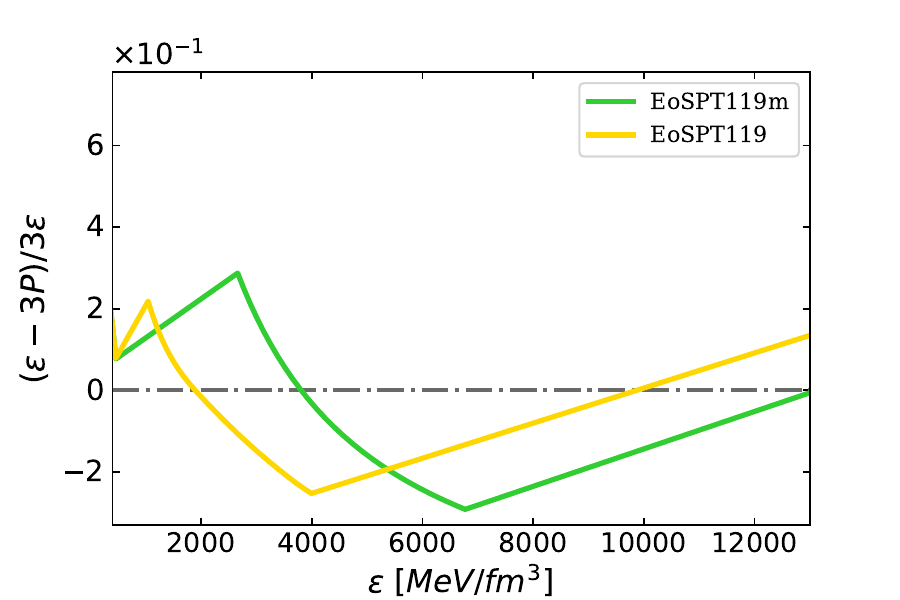}  \includegraphics[width=0.45\textwidth]{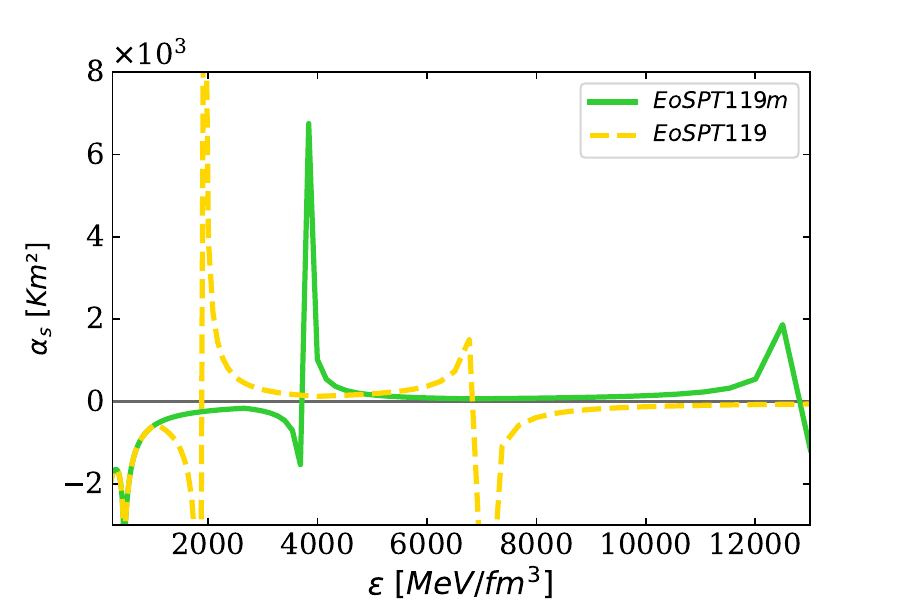}\\     \caption{{\bf Top:} Trace anomaly ($\varepsilon-3P)/3\varepsilon$ as a function of energy density for
 EoSPT119m (green curve in Figure \ref{fig:EoS_MvR}) and EoSPT119 (yellow curve in Figure \ref{fig:EoS_MvR_typeIa}). {\bf \textbf{Bottom}:} Values of the singular parameter $\alpha_s$ for the same EoSs.   
}
      \label{fig:AnomalyPT119}
  \end{figure}

The category II twin-stars  are obtained here in GR from EoS1 at $P=119$ MeV/fm$^3$ with  any jumps $425 < \Delta\varepsilon \leq 720$  MeV/fm$^3$, but for $\Delta\varepsilon \geq 720$  MeV/fm$^3$ only one twin star configuration is produced (violet line of Fig. \ref{fig:PT_typeI_II}) for $\Delta \varepsilon >720$ MeV/fm$^3$ (gray line), representing the largest jump allowed by pQCD constraints for this PT. We use the thi the twin-star category II is  here represented by the EoSPT119m and this enormous jump, corresponding  now to the green curve in Figure \ref{fig:EoS_MvR}, as an example of extreme jump.

However, in ETG, for all the jumps considered in this EoSPT119m, for $\Delta \varepsilon >720$ MeV/fm$^3$, twin-stars are obtained applying negative $\alpha$, in the same way than in the case Ia. 

The large jump leads to a vast range of negative trace ($\Delta\varepsilon_{neg}=8500$ MeV/fm$^3$), with the minimum occurring at the beginning of second phase transition, which can be seen in the curve green at the top of Figure \ref{fig:AnomalyPT119}. The plot at the bottom of Figure \ref{fig:AnomalyPT119} displays the values of the singular parameter $\alpha_s$, showing a maximum value for negative values at $\alpha_s=-165$ km$^2$ at the end of the long first PT at $P=119$ MeV/fm$^-3$.   

\begin{figure*} 
  \centering
  \advance\leftskip-0.7cm
 \subfloat{\includegraphics[width=.48\linewidth]{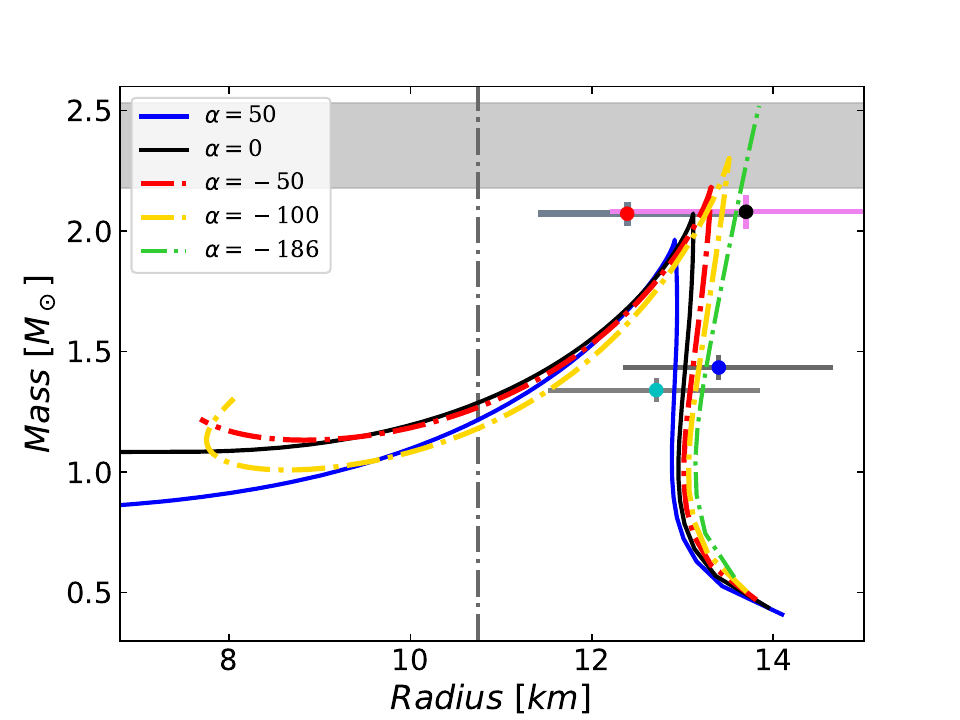}\quad\includegraphics[width=.54\linewidth]{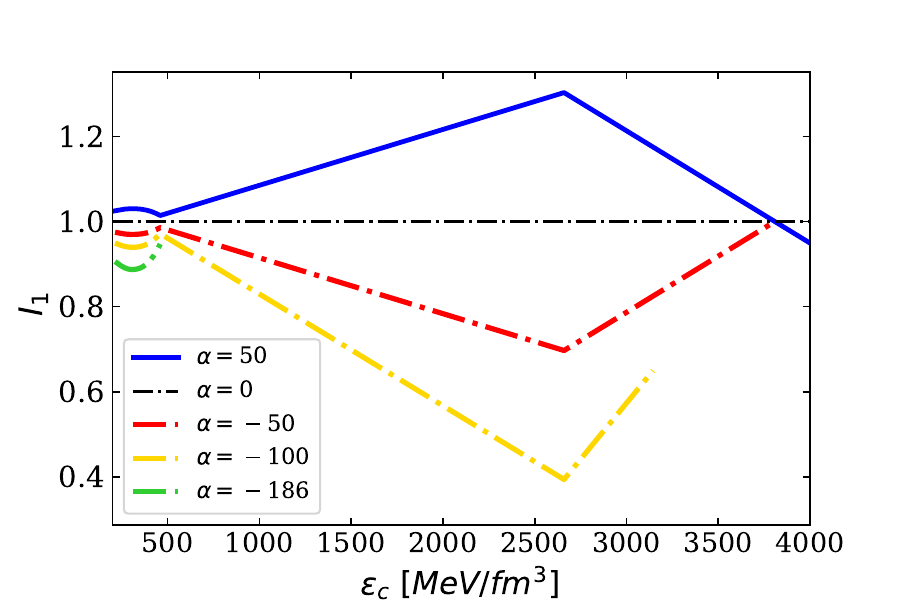}}\\[\baselineskip]
  \subfloat{\includegraphics[width=.515\linewidth]{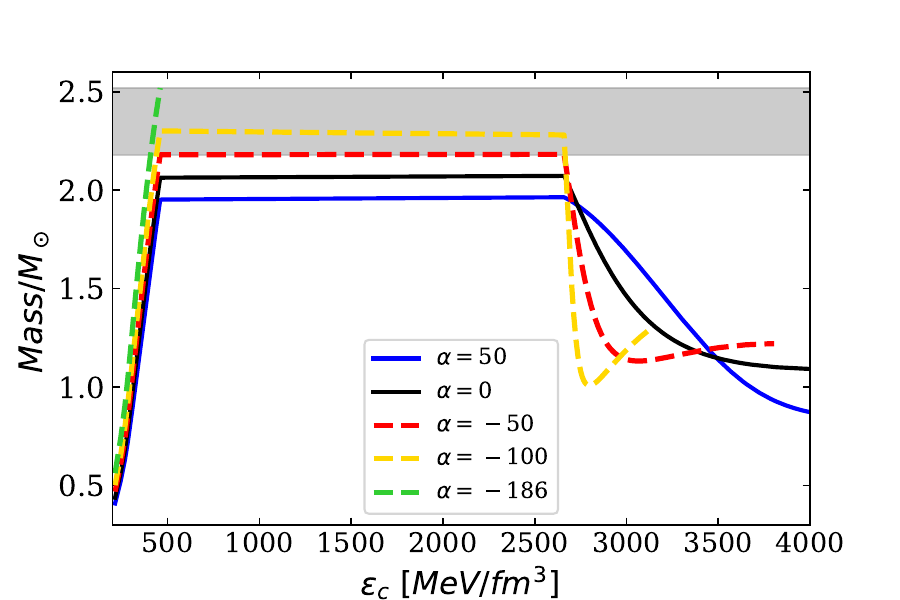}\quad\includegraphics[width=.515\linewidth]{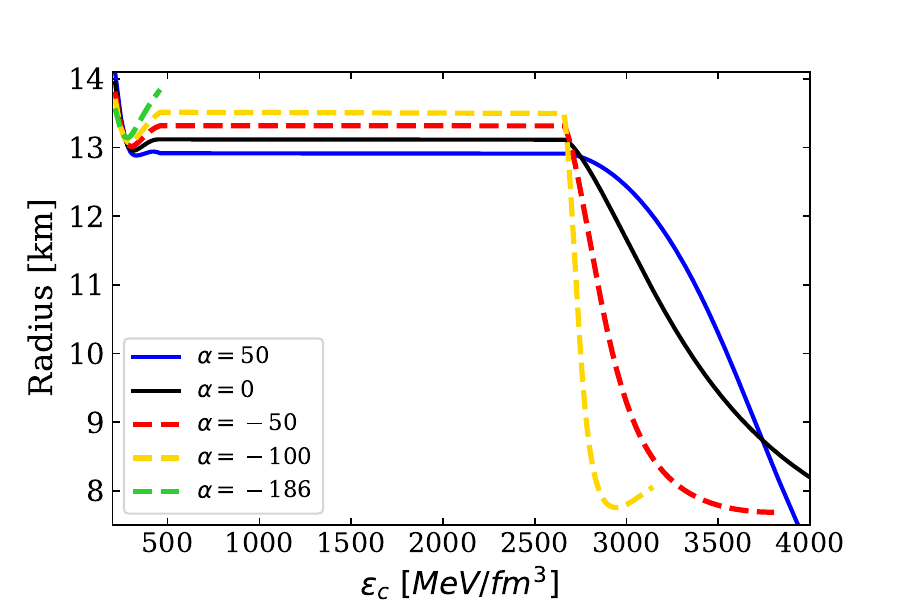}}
\caption{{\bf Top:} Mass-Radius diagrams (left) in GR (black solid line) and ETG (coloured lines) for the  EoSPT119 (green)  of Figure \ref{fig:EoS_MvR} as an example of Category II). The grayish shaded area contains the  lower and  upper mass bound (2.17 and 2.52 $M_\odot$, respectively) and the vertical dash-dotted line indicates the lower radius bound (10.8 km). Black and red circles indicate  measurements by  NICER of PSR J0740+6620 from \cite{Miller:2021qha} and \cite{Riley:2021pdl}, respectively, with error bars at 1 sigma confidence; blue and cyan circles indicate measurements by  NICER of PSR J0030+0451 from  \cite{Miller:2019cac} and \cite{Riley:2019yda}, respectively at 1 sigma confidence. On the right, $\mathcal{I}_1$ against the central energy density for different values of $\alpha$. {\bf Bottom:}  Mass (left) and Radius (right) against the central energy density for green EoS for different values of $\alpha$. }    
    \label{fig:MvREoSPT119max}
\end{figure*}

Permissible values of negative $\alpha$ should be greater than $\alpha_s=-165$ km$^2$ to cover the whole range of energy densities during this PT. However, such small values of alpha $-150\leq \alpha \leq  -13$ km$^2$  provide instabilities after the PT due to the negative traces produced by the necessary stiffness of the EoS. Figure \ref{fig:MvREoSPT119max} displays the results in ETG, showing (at the top) that the maximum mass of the grey shaded area  of the plot at the top left is reached with $\alpha=-186$ km$^2$, which fails before PT, since this value is lower than $\alpha_s$ and, consequently, no twin-star is generated.  Considering the allowed range of $\alpha$ across the entire range of PT, the maximum mass achievable is $M=2.45$ $M_\odot$ for $\alpha=-160$ km$^2$,not generating twin stars that are obtained only for $-140 \leq \alpha \leq 0 $ km$^2$. The lower mass bound is achieved for $\alpha=-50$ km$^2$.

The plot at the top left in Figure \ref{fig:MvREoSPT119max} displays the behavior of $\mathcal{I}_1$ until the energy density where the code becomes unstable or diverges. For negative $\alpha$, the decrease of $\mathcal{I}_1$ results in a faster drop of both mass and radius than in GR. Subsequent stiffening of the EoS leads to an increase in $\mathcal{I}_1$, causing the growth of mass and radius. Conversely, for positive $\alpha$, the opposite occurs.

The plots of mass and radius against energy density in Figure \ref{fig:MvREoSPT119max} show a different behavior than the corresponding ones for type I, due to the long phase transition and the large decrease in $\mathcal{I}_1$. After PT, for negative alphas, there is a greater decrease in mass and radius than in GR. For positive alphas, the behavior is the opposite.

In both GR and ETG, the second stable branch has too small radii not even compatible with the information from HESS J1731-347 \cite{2022NatAs...6.1444D}.

In summary, the allowed range of $\alpha$ dictated by this EoS is $\ -186 \leq \alpha < 50$ km$^2$, but the allowed range covering all the PT range and yielding twin stars is $\ -130 \leq \alpha <0$ km$^2$. 

\subsubsection{Category III}

   \begin{figure}
      \centering \includegraphics[width=0.45\textwidth]{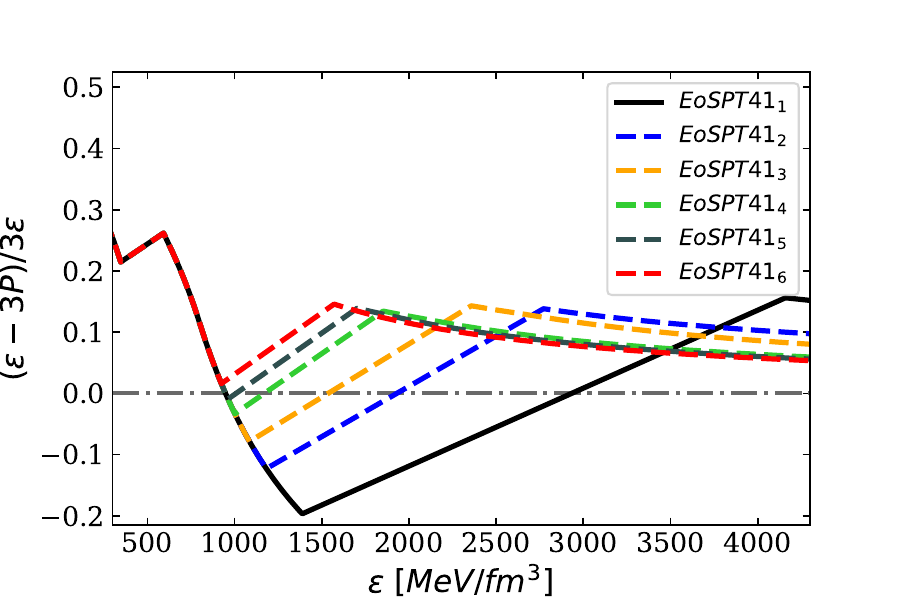}  \includegraphics[width=0.45\textwidth]{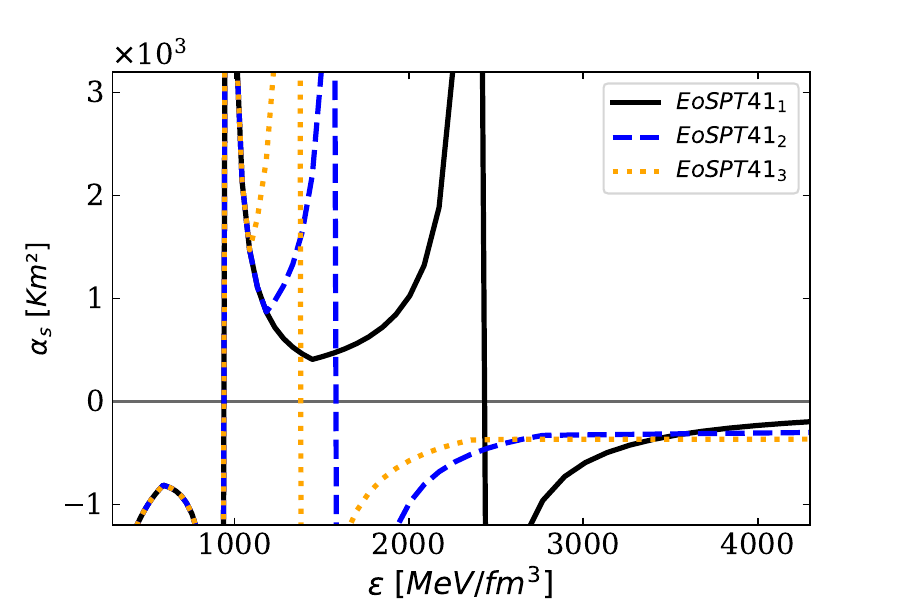}
      \includegraphics[width=0.45\textwidth]{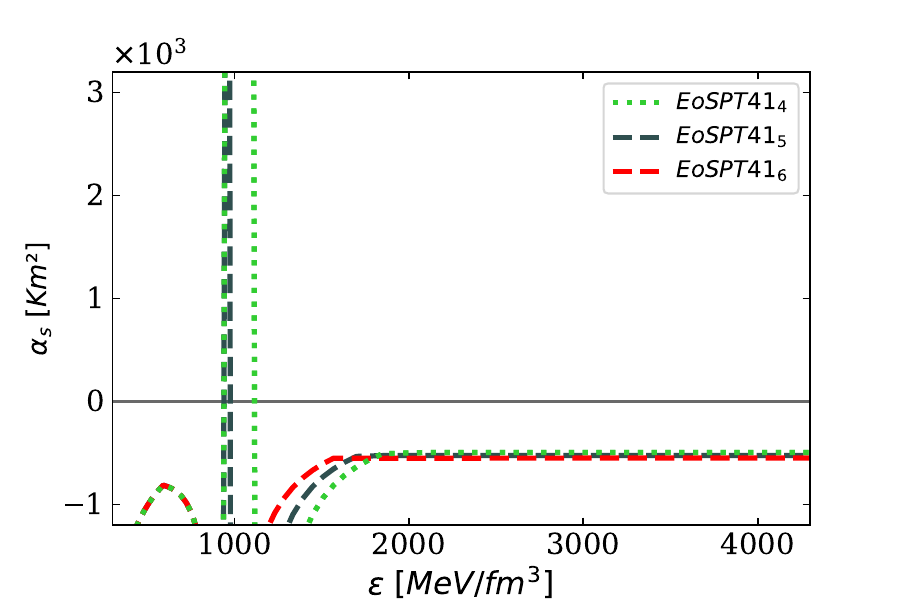}\\
      \caption{{\bf Top:} Trace anomaly ($\varepsilon-3P)/3\varepsilon$ as a function of energy density for the 6 EoS of Figure \ref{fig:EoS_PT41} and in the same colours. {\bf \textbf{Middle}:} Values of the singular parameter $\alpha_s$ for the black, blue and orange EoS, with decreasing range of negative trace, respectively. {\bf \textbf{Bottom}:} Values of the singular parameter $\alpha_s$ for the green, dark green  and red EoS, with decreasing range of negative trace, for the first two and positive trace for red EoS.    
}
      \label{fig:AnomalyPT41}
  \end{figure}

\begin{table}
    \centering
    \begin{tabular}{ccccc}
    \hline
       EoS & $(\Delta\varepsilon)_{neg}$  & $(\varepsilon-3P)_{min}$ &$\varepsilon_{min}$   &  $\alpha_{min}$ \\
      Twin type& MeV/fm$^3$& MeV/fm$^3$&  MeV/fm$^3$ &km$^2$ \\ \hline \hline
        I & 793 & -528 &1334 & -85 \\
        Ia & 4848 & -3030 & 3994 &-30  \\ \hline
       II & 8664 & -5930 & 6773 &-13\\ \hline
       III$_1$ & 1407 & -928 &1446 & -58 \\
       III$_2$& 605 & -436 & 1182 & -93\\
       III$_3$& 372 & -257& 1086 &-119\\
       III$_4$& 124 & -101& 1002 & -158\\
       III$_5$& 27 & -15& 962 &-187\\
       III$_6$& 0 & +44& 924 &-227\\ \hline
       IV& 224 & -184 &887 &-152\\
    \end{tabular}
    \caption{Minimum values of $\alpha$ (km$^2$) associated to the minimum value of the trace $(\varepsilon-3P)_{min}$, at the energy density $\varepsilon_{min}$,  within the corresponding range of negative traces ($\Delta\varepsilon_{neg}$), for the  twin stars EoS of Figure \ref{fig:EoS_MvR} (types I, II and IV), Figure \ref{fig:EoS_MvR_typeIa} (type Ia EoS) and Figure \ref{fig:EoS_PT41} (Type III$_1$ to III$_6$ EoS).}
    \label{tab:ETG}
\end{table}

The twin-star category III is represented by EoSPT41, corresponding to the red curve in Figure \ref{fig:EoS_MvR}, and EOSPT41$_1$ of Figure \ref{fig:EoS_PT41} (black line). Sequentially, we constructed five smoother EoSs with a progressive decrease in maximum mass and radius in GR.

In ETG, since these EoSs with PT at $P=41$ are soft, negative alphas are needed to reach both the lower and upper mass limits estimated for PSR J0952-0607. Here we adopt this family of 6 EoS to address the range $\Delta\varepsilon_{neg}$ as well as the values $(\varepsilon-3P)_{min}$ of the negative traces tolerated by ETG for negative values of $\alpha$.

 Figure \ref{fig:AnomalyPT41} (top) displays the trace anomaly for the six EoSPT41 (subindex 1 to 6) of Figure \ref{fig:EoS_PT41} and in the same colours, showing the decrease of $\Delta\varepsilon_{neg}$ and the increase of  $(\varepsilon-3P)_{min}$ from EoS 1 to 6. The middle and bottom plots displays the values of $\alpha_s$ for EoS 1 to 3 and 4 to 6, respectively, in the same range of energy density. The first singularity of $\alpha_s$ is common for the the five EoS  at $\varepsilon=948$ MeV/fm$^3$. As the range  $\Delta\varepsilon_{neg}$ is decreased, lower maximum (negative) $\alpha_s$ are obtained at smaller energy densities. 

Table 5 summarises the values of $\alpha_{min}$ of these 6 EoS for each range $\Delta\varepsilon_{neg}$ and minimum trace values $(\varepsilon-3P)_{min}$,  showing the decreasing of $\alpha_{min}$ as increasing the values of $(\varepsilon-3P)_{min}$. However, for $\alpha < -200$ km$^2$ the code becomes unstable or/and diverges for positive values of the trace for EoS 1 to 5. It is also worthwhile mentioning that, for EoS 6, with positive trace over the whole range of energy densities inside NS, $\alpha_{min}=-225$ km$^2$ ($\beta \approx -4 \times 10^6$ m$^2$).

\begin{figure*} 
  \centering
  \advance\leftskip-0.7cm
 \subfloat{\includegraphics[width=.48\linewidth]{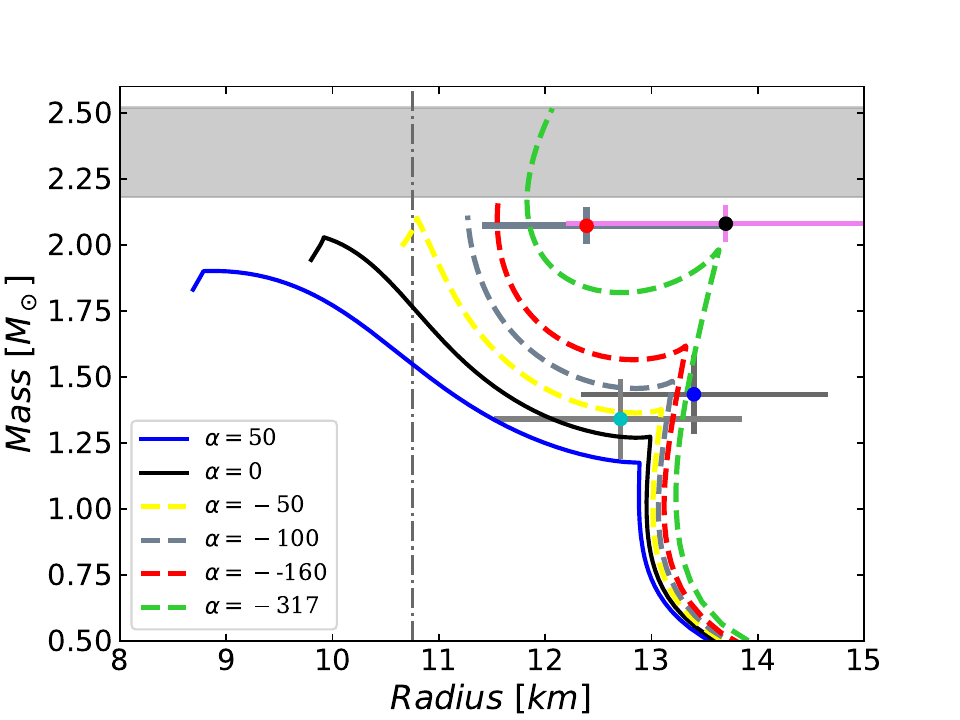}\quad\includegraphics[width=.54\linewidth]{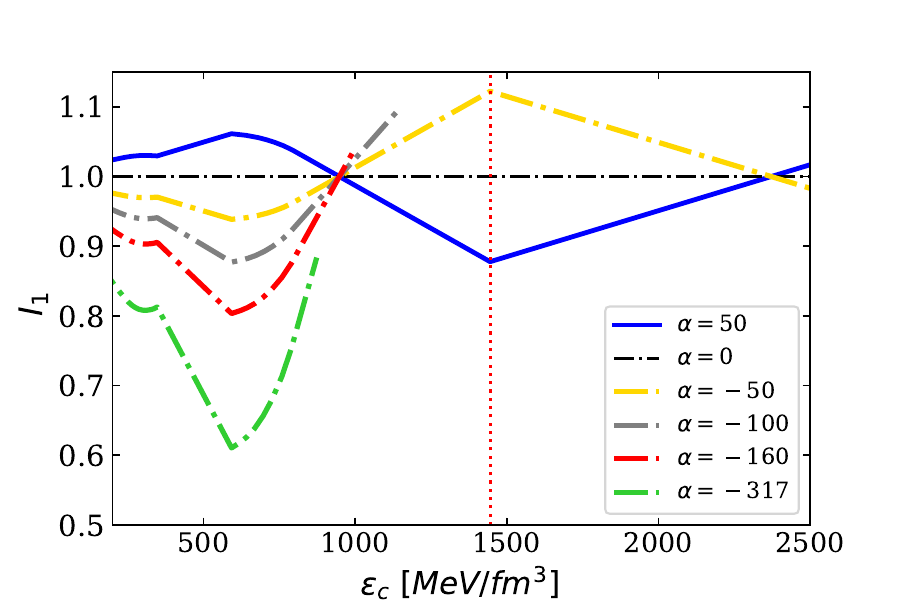}}\\[\baselineskip]
  \subfloat{\includegraphics[width=.48\linewidth]{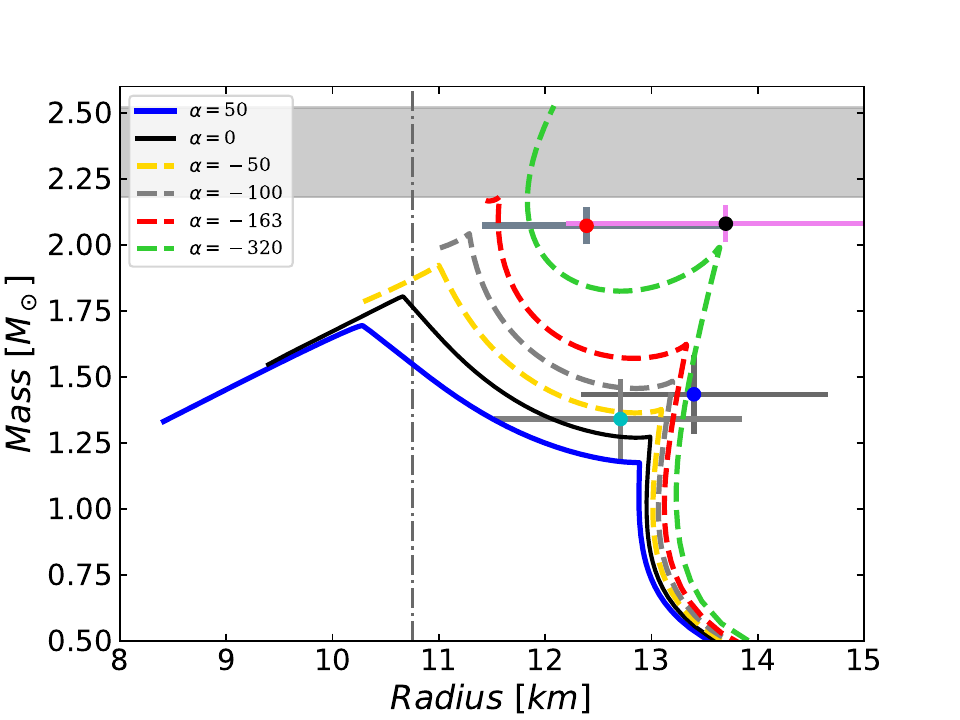}\quad\includegraphics[width=.54\linewidth]
  {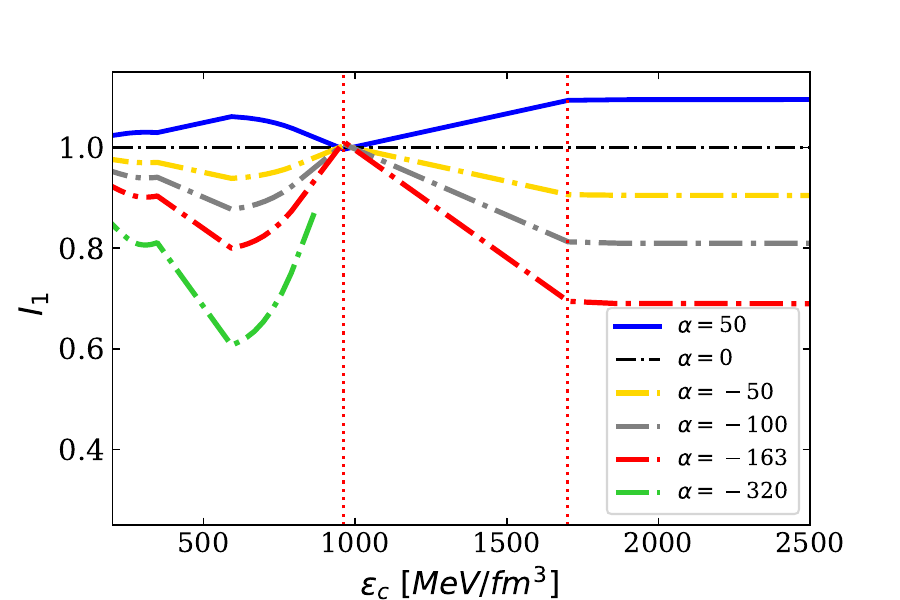}}\\[\baselineskip]
  \subfloat{\includegraphics[width=.48\linewidth]{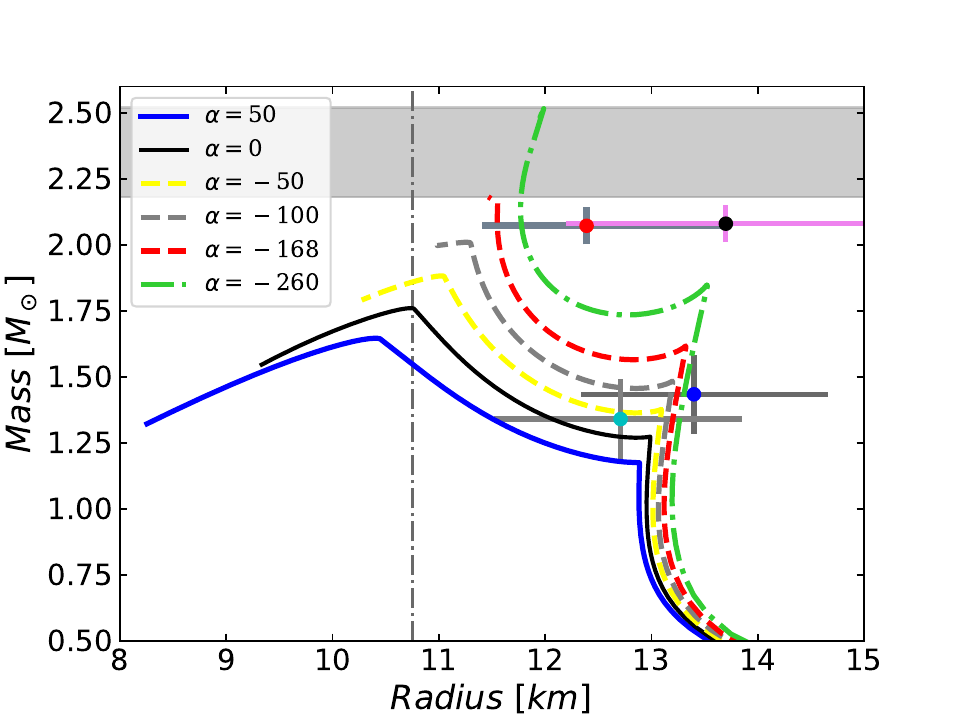}\quad\includegraphics[width=.54\linewidth]{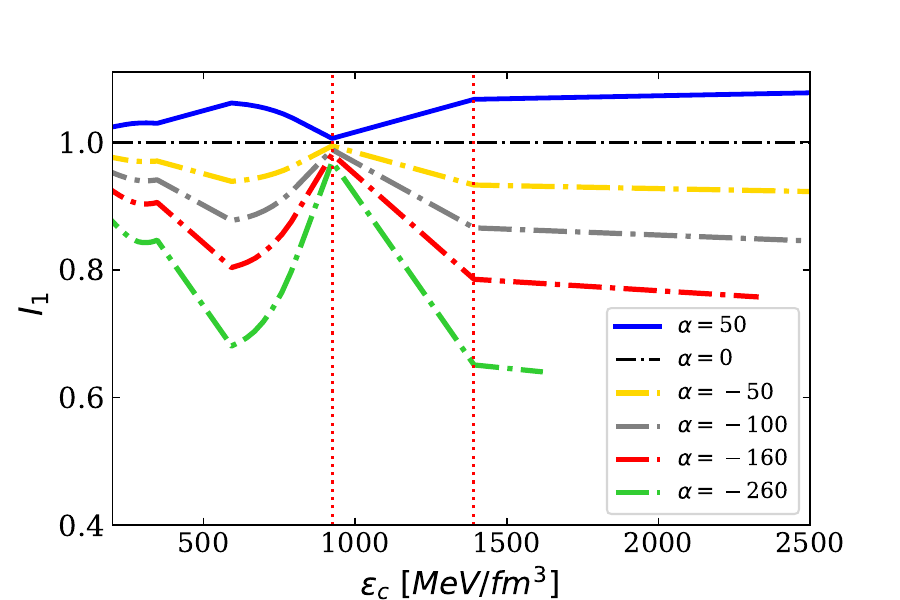}}
\caption{{\bf Top:} Mass-Radius diagrams (left)  in GR (black solid line) and ETG (coloured lines) for the  EoSPT41$_1$ (darkgreen)  of Figure \ref{fig:EoS_PT41} as an example of Category III with long  $\Delta\varepsilon_{neg}$. The grayish shaded area contains the lower and  upper mass bound (2.17 and 2.52 $M_\odot$, respectively) and the vertical dash-dotted line indicates the lower radius bound (10.8 km). Black and red circles indicate  measurements by  NICER of PSR J0740+6620 from \cite{Miller:2021qha} and \cite{Riley:2021pdl}, respectively, with error bars at 1 sigma confidence; blue and cyan circles indicate measurements by  NICER of PSR J0030+0451 from  \cite{Miller:2019cac} and \cite{Riley:2019yda}, respectively at 1 sigma confidence. On the right, $\mathcal{I}_1$ against energy density with the same colours. 
{\bf Second row:}  M-R (left) and $\mathcal{I}_1$  against the central energy density (right) for the  EoSPT41$_5$ gray EoS for different values of $\alpha$, as example of very small density range of negative trace. {\bf Bottom:}  M-R (left) and $\mathcal{I}_1$  against the central energy density (right) for the  EoSPT41$_6$ gred EoS for different values of $\alpha$, as example of positive trace over the whole density range.}    
    \label{fig:MvREoSPT41_1}
\end{figure*}

The M-R diagrams of three of the six EoS in Fig. \ref{fig:EoS_PT41} are shown in the plots on the left of Figure \ref{fig:MvREoSPT41_1}: EoSPT41$_1$ (black) at the top, EoSPT41$_5$ (grey) in the centre and EoSPT41$_6$ (red) at the bottom. The plots on the right of Figure \ref{fig:MvREoSPT41_1} show the behaviour of $\mathcal{I}_1$ up to the energy density where the code becomes unstable or diverges for the same EoSs. The lower level of the shaded mass band ($M=2.18 M_\odot$) is reached for very similar values of $\alpha$ (from -163 to -168 km$_2$) in all EoSs, but with different negative trace tolerances for each. Thus, small negative trace values ($\Delta\varepsilon_{neg} = 48$ MeV/fm$^3$) are tolerated for EoSPT41$_1$, while for EoSPT41$_5$ and EoSPT41$_6$, the full range of $\Delta\varepsilon_{neg}$ is completed. In order to reach the upper level of the shaded mass band ($M=2.52 M_\odot$), very small values of $\alpha$ are required, which fail at the beginning of $\Delta\varepsilon_{neg}$.

The mass and radius gap in the unstable branch is larger than in GR, and this difference increases with decreasing alpha.  Moreover, in the subsequent stable branch, the mass grows faster than in GR and the radius remains almost constant or grows, as opposed to its decrease in GR. Both behaviours cause a peculiar M-R diagram of the twin-stars in ETG.

The three M-R diagrams in Figure \ref{fig:MvREoSPT41_1} show that, unlike in GR, these twin-stars fulfill the NICER radius requirements.

\subsubsection{Category IV}
   \begin{figure}
      \centering      \includegraphics[width=0.45\textwidth]{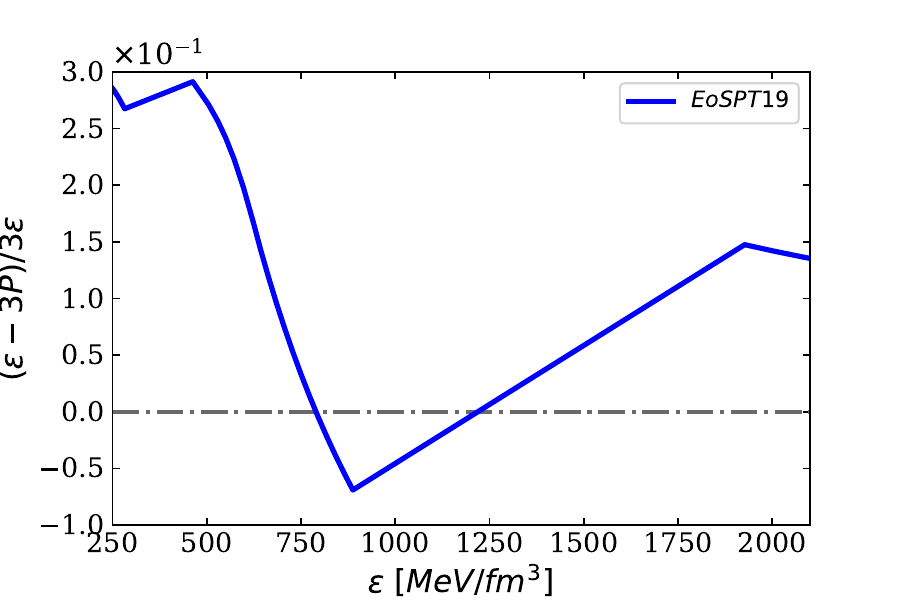}  \includegraphics[width=0.45\textwidth]{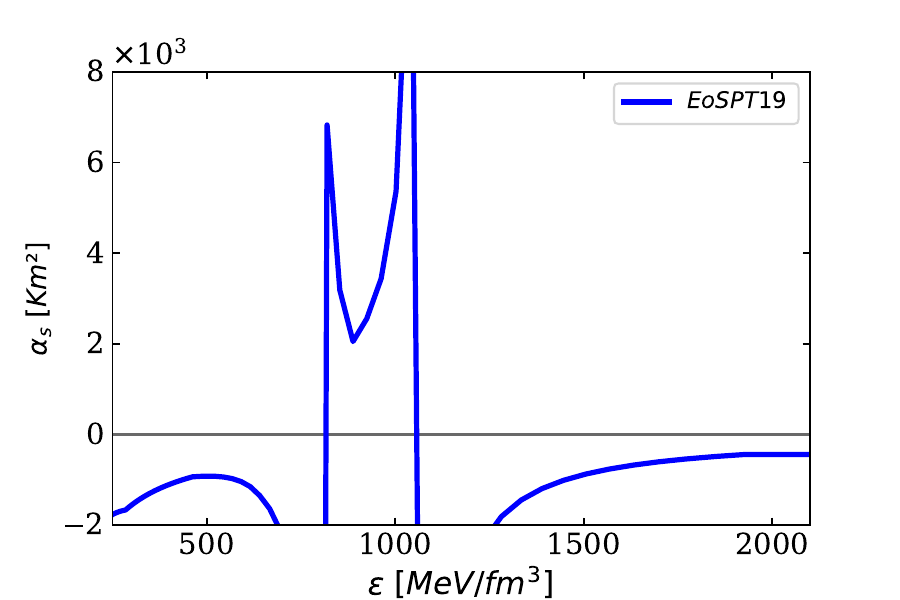}\\     \caption{{\bf Top:} Trace anomaly ($\varepsilon-3P)/3\varepsilon$ as a function of energy density for
 EoSPT19 (blue curve in Figure \ref{fig:EoS_MvR}). {\bf \textbf{Bottom}:} Values of the singular parameter $\alpha_s$ for the same EoSs.   
}
      \label{fig:AnomalyPT19}
  \end{figure}

This category is represented by the blue line in Figure \ref{fig:EoS_MvR}, as an example of EoS that generates an almost twin-star in GR (the unstable branch is very small), but yields twin-stars in ETG, similar to example 1a. 
 
Figure \ref{fig:AnomalyPT19} displays trace anomaly (top) and $\alpha_s$ (bottom), showing that maximum negative $\alpha_s=-440$ km$^2$ after the second PT. However, the minimum value of the trace at $\varepsilon_{min}=1043$ MeV/fm$^3$ limits the value of alpha to $\alpha > \alpha_{min}=-152$ km$^2$

\begin{figure*} 
  \centering
  \advance\leftskip-0.7cm
 \subfloat{\includegraphics[width=.50\linewidth]{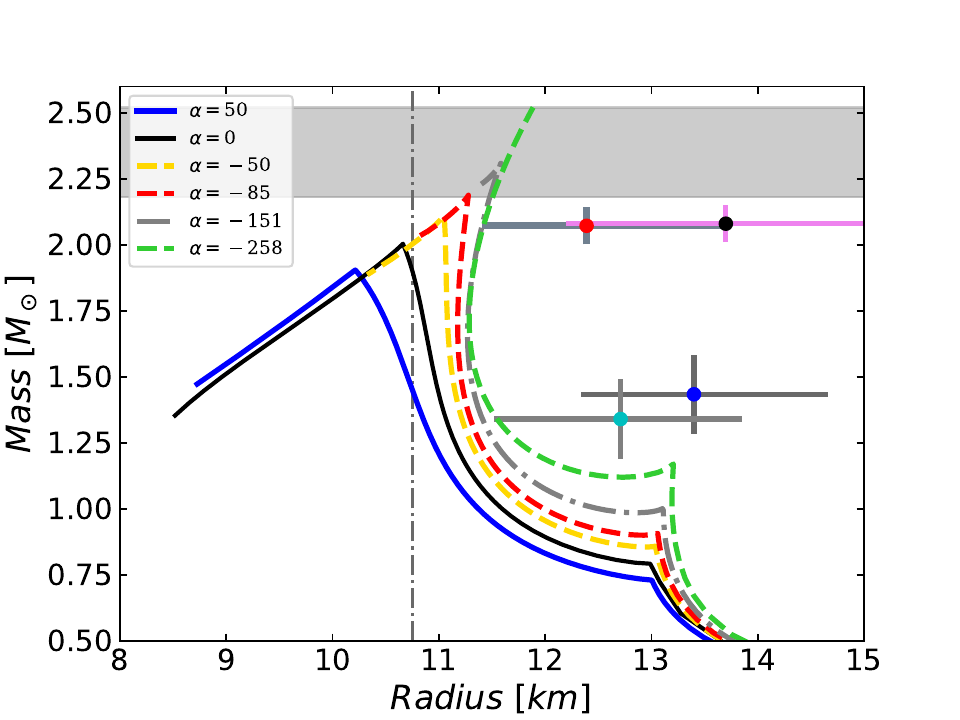}\quad\includegraphics[width=.53\linewidth]{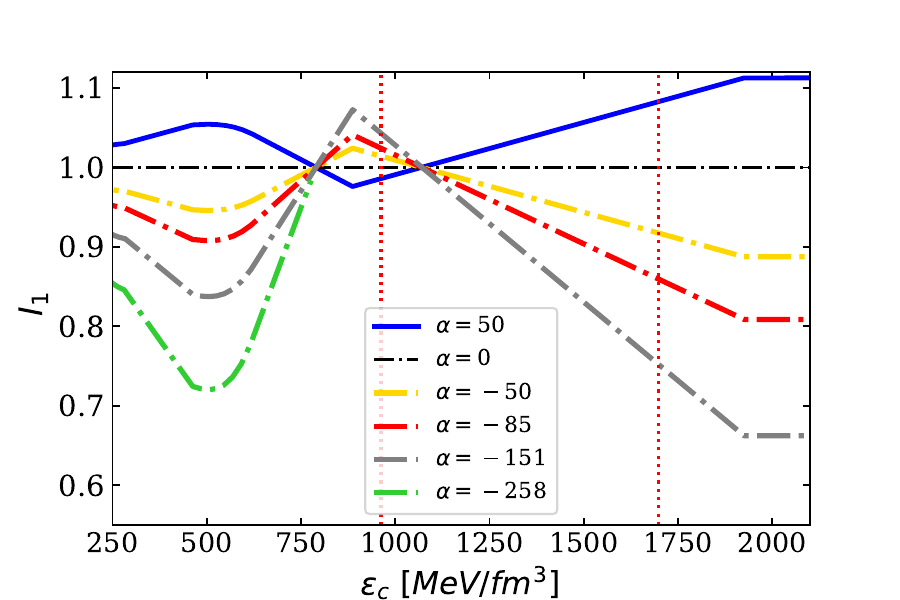}}\\[\baselineskip]
  \subfloat{\includegraphics[width=.515\linewidth]{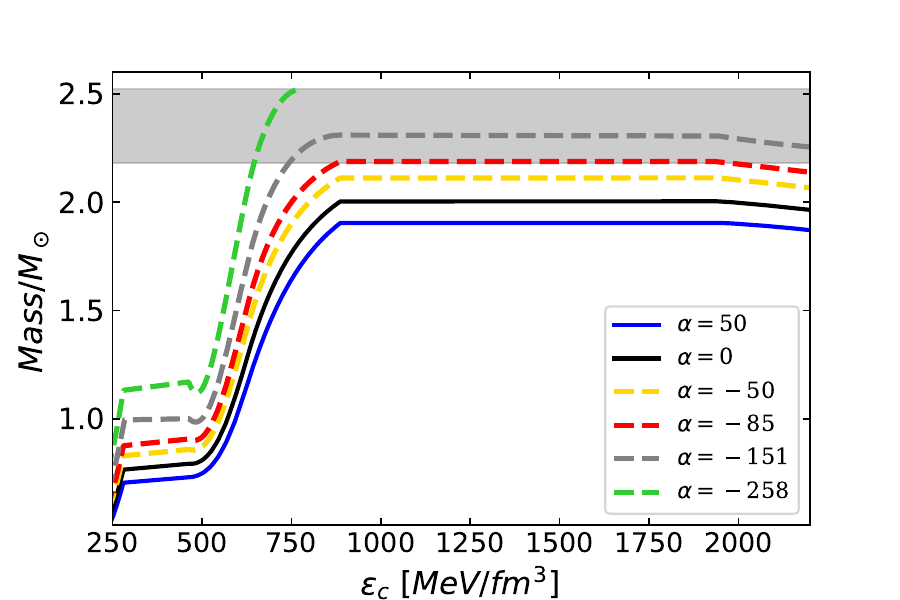}\quad\includegraphics[width=.515\linewidth]{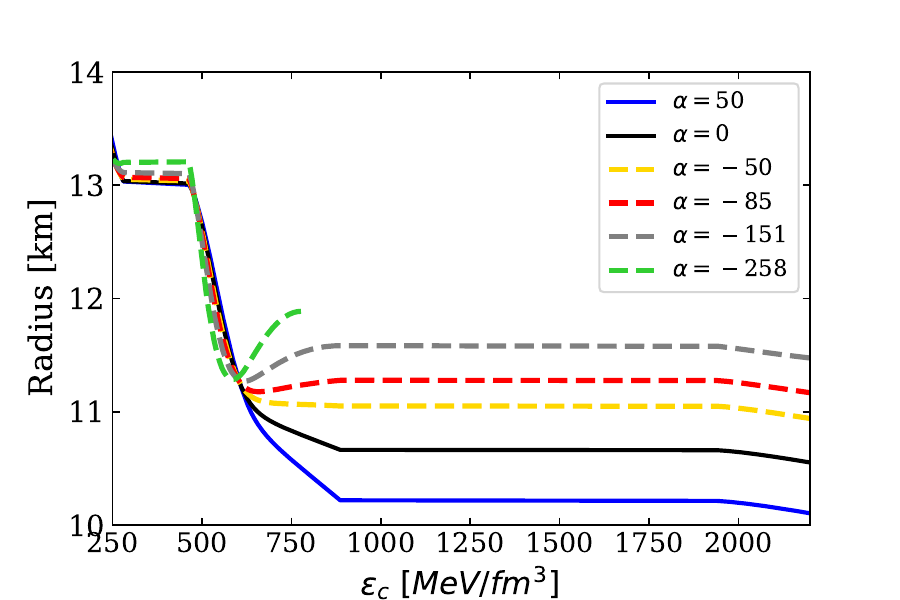}}
\caption{{\bf Top:} Mass-Radius diagrams (left) in GR (black solid line) and ETG (coloured lines) for the  EoSPT19 (red)  of Figure \ref{fig:EoS_MvR} as an example of Category IV with  with a small $\Delta\varepsilon_{neg}$. The grayish shaded area contains the  lower and  upper mass bound (2.17 and 2.52 $M_\odot$, respectively) and the vertical dash-dotted line indicates the lower radius bound (10.8 km). Black and red circles indicate  measurements by  NICER of PSR J0740+6620 from \cite{Miller:2021qha} and \cite{Riley:2021pdl}, respectively, with error bars at 1 sigma confidence; blue and cyan circles indicate measurements by  NICER of PSR J0030+0451 from  \cite{Miller:2019cac} and \cite{Riley:2019yda}, respectively at 1 sigma confidence. On the right, $\mathcal{I}_1$ against the central energy density for different values of $\alpha$.  {\bf Bottom:}  Mass (left) and Radius (right) against the central energy density for green EoS for different values of $\alpha$. }    
    \label{fig:MvREoSPT19}
\end{figure*}

Figure \ref{fig:MvREoSPT19} displays GR and ETG results aiming to reach the lower and upper mass bounds of PSR J0952-0607, as well as the radius bound from NICER. The top-left plot shows that $-258 \leq \alpha \leq-85$ km$^2$ is required to meet the mass bounds, although the lower bound of the radius (at 1$\sigma$) is achieved for $\alpha \leq -151$ km$^2$, reaching $M=2.31 M_\odot$ for $\alpha =-151$ km$^2$. As in the rest of EoS, the code yields instabilities for small positive traces for $\alpha <-200$ km$^2$, reaching the upper mass bound $M=2.52 M_\odot$ for $\alpha =-258$ km$^2$, which fails for a positive trace.

The top-right plot illustrates $\mathcal{I}_1$ against energy density for each $\alpha$. For negative alphas, this conformal parameter increases as the EoS stiffens up to the phase transition, but decreases when a PT begins, reaching the minimum value at the end of the PT. After the second PT, the EoS continues with a conformal slope until it enters pQCD, resulting in a small decrease of $\mathcal{I}_1$ if $c_s^2 <1/3$ and a practically constant value when $c_s^2 <1/3$. Furthermore, M-R diagrams show degeneracy for  $\alpha < -151$ km$^2$ and $\alpha=-258$ km$^2$ from $M>$ 1.5 $M_\odot$. The bottom-left plot shows mass against energy density, illustrating how ETG increases mass as well as the unstable branch when decreasing $\alpha$. The bottom-right plot shows radius against energy density, revealing a lower decrease when decreasing $\alpha$ and, for each $\alpha$, when decreasing $\mathcal{I}_1$. However, the radius increases as $\mathcal{I}_1$ grows up to the second PT begins, where the radius remains constant.

All twin-star generated for this EoS by negative alphas fulfill NICER requirements at 2$\sigma$ confidence, with larger radius increases between twins than in GR, while remaining smooth at low energy densities.

\section{Conclusions}\label{seccon}
In this investigation, we delved into a more detailed exploration of gravity-independent equations of state on neutron stars developed in \cite{LopeOter:2019pcq,Lope-Oter:2023urz}, particularly focusing on phase transitions and their interplay with the twin star problem.
The construction of these equations incorporated insights from NN and 3N chiral potentials for low-density scenarios, perturbative QCD for high-density regimes, and constraints from monotonicity and causality for intermediate densities. This approach aimed to uphold model independence while adhering to fundamental physics principles.

To achieve our goals, we selected the intermediate EoS1 from Figure \ref{fig:3EoS_Twin} due to its versatility in generating all four types of twin stars while aligning with the chiral band for $n<0.32$ fm$^{-3}$ \cite{Drischler:2020hwi}. EoS1 also satisfied the tidal and radius constraints of GW170817 \cite{LIGOScientific:2018cki} and NICER \cite{Miller:2019cac,Riley:2019yda,Miller:2021qha,Riley:2021pdl}. 

Our focus centered on the four twin star categories, each originating from four initial phase transitions in EoS1 (black curve in Figure \ref{fig:EoS_MvR}). These transitions occurred at pressures $P=139$ MeV/fm$^3$ (yellow) for type I, $P=119$ MeV/fm$^3$ (red) for type II, $P=41$ MeV/fm$^3$ (green) for type III, and $P=19$ MeV/fm$^3$ (blue) for type IV. All these phase transitions emanated from the stiff hadronic branch of EoS1.

In the realm of General Relativity, Type I and Type II twin stars pose challenges, requiring exceptionally high slopes post-phase transition for stability. Achieving this involves careful chemical potential control, smoothing the equation of state predictably through prolonged transitions with a maximal slope of $c_s^2=1$.

Type I twins are found in three equations of state, exhibiting non-extensive jumps at each phase transition. However, specific pressure ranges in certain equations of state yield no Type I twins. The mass gap for this category is $0.01< \Delta M <0.34 M_\odot$, with a latent heat range of $0.094 \leq L \leq 0.149$. Some results may be subject to NICER confirmation in General Relativity (GR).

Type II twins, confirmed within specific pressure ranges in certain equations of state, exhibit jumps exceeding the Seidov limit. This category provides a mass gap of $0.1< \Delta M <1.3 M_\odot$, with a latent heat spanning $0.12 \leq L \leq 0.22$. All these cases could potentially be confirmed by NICER in GR.

In contrast, Type III and Type IV twins emerge in all considered equations of state for jumps $\Delta\varepsilon > \Delta\varepsilon_{crit}$, with variable lengths based on the target mass $M_2$. Both types require stiffening the equation of state post-phase transition to achieve $M_2\geq 2M_\odot$, leading to a density range with negative traces.
To balance these considerations, the mass gap between stable branches is reduced, and the slope of the equation of state post-phase transition is gradually accelerated.
Type III involves a phase transition with specific characteristics. In GR, this EoS fails to meet NICER requirements. Type IV exhibits distinct stable branches in a mass-radius diagram.

In the context of Extended Theories of Gravity (ETG), the trace of the energy-momentum tensor appearing in the modifying term $\mathcal{I}_1$ exhibits a significant interplay with the parameter $\alpha$. For positive $\alpha$, rigidity in the equation of state leads to a decrease in $\mathcal{I}_1$, whereas for negative $\alpha$, the trend is reversed. During phase transitions, the behavior changes, reaching a maximum at the end of the phase transition for positive alphas and a minimum for negative alphas, highlighting the sensitivity of $\mathcal{I}_1$ to the underlying physics.

In ETG, differences from General Relativity arise across five twin-star categories: I, Ia, II, III, and IV.
\begin{itemize}
 \item     Type I and Ia Twin Stars:
        In ETG, negative values of $\alpha$ are required for type I and Ia twin stars, presenting a larger mass gap compared to GR.
        Positive $\alpha$ in ETG does not produce twin stars in these categories. NICER confirmation for these cases in GR is subject to constraints, with some results rejected. Note that type Ia exists only in ETG according to our findings.
 \item     Type II Twin Stars:
        Extreme jumps in energy density ($\Delta\varepsilon > 720$ MeV/fm$^3$) lead to category II twin stars in GR, represented by EoSPT119.
        In ETG, negative $\alpha$ values are essential for twin-star configurations, showing notable sensitivity to the density interval with negative traces. NICER requirements may still be met.
 \item     Type III Twin Stars:
        Category III twin stars in ETG (EoSPT41) exhibit different behavior than in GR, with a sensitivity to the density interval with negative traces.
        Negative $\alpha$ values are crucial in ETG for twin-star existence, and significant oscillations and divergences occur, impacting the M-R diagram. NICER requirements can be fulfilled at 1$\sigma$ confidence for specific cases.
 \item     Type IV Twin Stars:
        ETG introduces differences in category IV twin stars compared to GR, generating twin-star configurations in ETG not present in GR.
        Negative $\alpha$ values are needed for these twin stars in ETG, and they can meet NICER requirements at 2$\sigma$ confidence.
\end{itemize}

Overall, the analysis highlights the sensitivity of twin-star solutions in ETG to the parameter $\alpha$. In ETG, positive $\alpha$ increases both mass and radius, while negative $\alpha$ decreases both mass and radius. To determine $\alpha$ for each EoS, knowledge of the maximum $\alpha_{\text{s}}$ for positive $\alpha$ and the minimum $\alpha_{\text{s}}$ for negative $\alpha$ is required. In EoS with phase transitions, the selection of $\alpha$ is determined by the minimum/maximum value at the end of the last phase transition (inside central values of the star). Since most generated twin-stars reach maximum masses slightly above $2 M_\odot$ and, in general, radii below the 1$\sigma$ NICER limit for these masses, negative alphas are needed to increase masses and radii.

The use of negative alphas (compatible with $\alpha_{\text{s}}$) generates instabilities for energy densities with negative or near-zero trace, and these instabilities grow with decreasing $\alpha$. For negative alphas, the conformal factor $\mathcal{I}_1$ increases while the EoS is stiffening but decreases significantly during a phase transition, depending on $\alpha$. Thus, ETG accentuates the unstable branch in mass-radius diagrams, depending on the jump of the phase transition and the $\alpha$ value. For positive $\alpha$, the behavior is the opposite.

One should be also careful with statements such as "future radius measurement of the NICER mission have the potential to
reveal the existence of a strong phase transition in dense neutron star matter by confirming the
existence of so called twin stars" and that "NICER serving
as a smoking gun for a strong phase transition in neutron star matter" \cite{christian2022confirming}. We could see that while in GR there can exist EoS which does not provide twin star phenomena while in ETG, with the use of the same EoS, one can have it (see the new class Ia discussed). 

Let us also briefly discuss the general problem with the negative trace of the energy momentum tensor. Note that in both theories the trace is proportional to the curvature scalar:
\begin{equation}
    R=-\kappa^2 T,\;\;\;\;\;   \mathcal R=-\kappa^2 T.
\end{equation}
Because of that fact, when the trace changes the sign, the curvature of spacetime also does, changing the attractive feature of gravity to the repulsive one. Moreover, the negative trace violates the strong energy condition $T>0$, and although in some EoS one permits this behaviour for exotic matter (see, e.g. discussion in \cite{podkowka2018trace}), it should be treated carefully, especially, as discussed in the paper, in the case of ETG.

\section*{Acknowledgements}
 AW acknowledges financial support from MICINN (Spain) {\it Ayuda Juan de la Cierva - incorporaci\'on} 2020 No. IJC2020-044751-I.

 This preprint is issued with number IPARCOS-UCM24-006.

\bibliographystyle{apsrev4-1}
\bibliography{biblio}

\end{document}